\documentclass[12pt, a4paper]{article}

\pdfoutput=1

\usepackage{amsmath}
\usepackage{amsfonts}
\usepackage{amssymb}
\usepackage{cite}
\usepackage{multicol,multirow}
\usepackage{graphicx, physics, subcaption}
\usepackage{graphics, epsfig}
\usepackage{epsf}
\usepackage{epstopdf}
\usepackage[usenames,dvipsnames]{color}

\usepackage{colortbl}
\definecolor{summersky}{cmyk}{0.71,0.33,0,0.14}
\definecolor{flamingo}{cmyk}{0,0.51,0.71,0.14}
\definecolor{rp}{cmyk}{0.2, 1, 0.6, 0}
\definecolor{pacificblue}{cmyk}{0.95,0.3,0, 0.19}
\definecolor{gray60}{cmyk}{0.4,0.4,0,0.8}

\numberwithin{equation}{section}

\usepackage[colorlinks,bookmarks=false, hypertexnames=true]{hyperref}
\hypersetup{
	breaklinks=true,
	pdfstartview={FitH},    
	colorlinks=true,       
	linkcolor=blue,          
	citecolor=red,        
	filecolor=magenta,      
	urlcolor=red,           
	anchorcolor=green,      
	linktocpage=true
}

\newcommand{\nc}{\newcommand}

\nc{\ba}{\begin{eqnarray}}
\nc{\ea}{\end{eqnarray}}

\nc{\calR}{{\cal{R}}}
\nc{\calP}{{\cal{P}}}
\nc{\cN}{ {\cal{N}} }

\def\bfk{{\bf k}}

\begin{document}

\def\thefootnote{\fnsymbol{footnote}}

\begin{center}

{\bf  Upper Bounds on the Mass of Fundamental Fields from Primordial Universe
}
\\[0.5cm]

{
Hassan Firouzjahi\footnote{firouz@ipm.ir}
}
\\[0.5cm]
 
 {\small \textit{School of Astronomy, Institute for Research in Fundamental Sciences (IPM) \\ P.~O.~Box 19395-5531, Tehran, Iran
}}\\

\end{center}

\vspace{.3cm}
\hrule

\begin{abstract}
We study the fluctuations in the vacuum zero point energy associated to quantum fields and their statistical distributions during inflation. It is shown that the perturbations in the vacuum zero point energy  have large amplitudes which are highly non-Gaussian. The effects of vacuum zero point fluctuations can be interpreted as the loop corrections in primordial power spectrum and bispectrum.  Requiring that the primordial curvature perturbation to remain nearly  Gaussian and the loop corrections to be under perturbative control   impose strong upper bounds on the mass of fundamental fields during inflation. 
These bounds  depend on the hierarchy of the masses in the theory such as whether or not the masses are at the similar orders. 
 While the mass of the heaviest field in the hierarchy may not be constrained but it is shown that  a combination of the masses  of the  fields can not be much heavier than  the Hubble scale during inflation, otherwise their vacuum zero point fluctuations  induce large non-Gaussianities in primordial perturbations.  Considering the observational upper bound on tensor to scalar ratio, we conclude that this combined mass scale 
is lighter than $10^{14}$ GeV.

\end{abstract}


\section{Introduction}
\label{intro}

Vacuum zero point energy is a fundamental property of quantum mechanics, having its origin from the fact that the operators like position and momentum do not commute in quantum mechanics. The effects of vacuum zero point energy become more pronounced in quantum field theory  where particles and antiparticles can be created and annihilated continuously in vacua.  The reality  of vacuum zero point energy were confirmed in Casimir
effect  \cite{Casimir:1948dh, Lamoreaux:1996wh, Bordag:2001qi}. The roles of vacuum zero point energy become even more significant when one deals with  gravity. Based on Einstein field equation, any source of energy will act as a source of gravitation and the curvature of spacetime. 
Locally, the effects of the quantum vacuum zero point energy appears as a cosmological constant term in the Einstein field equation. Based on  equivalence principle, one expects the energy momentum tensor associated to the vacuum zero point fluctuations to be locally Lorentz invariant. Consequently,  the vacuum expectation values of  the pressure and the energy density are simply  related to each other as $\langle P_v\rangle =-\langle \rho_v\rangle$, where here and below the subscript  ``$v$" stands for vacuum.

The vacuum zero point energy associated to quantum perturbations of a fundamental field with mass $m$ is UV divergent. To regularize the quartic UV divergence, one may put a cutoff $\Lambda$, obtaining $\langle \rho_v\rangle \sim \Lambda^4$. Assuming that $\Lambda$ is given by a natural scale of the theory, such as the TeV scale of Standard Model (SM) of particle physics, one obtains the magnitude of vacuum zero point energy be roughly at the order  $(\mathrm{TeV})^4$. Alternatively, if one assumes $\Lambda$ to be at the order of Planck mass $M_P$, then the vacuum energy density becomes at the order  $M_P^4$. Of course, the trouble is that both of these predicted values are grossly in contradictions with observations. Indeed, various cosmological observations \cite{Planck:2018vyg, SupernovaCosmologyProject:1998vns, SupernovaSearchTeam:1998fmf} indicate that the  Universe is accelerating now with an unknown  source of energy density, the so-called dark energy, which is roughly at the order $(10^{-3} \mathrm{eV})^4$ which is vastly smaller than what one may naively obtain from basic quantum field theory analysis.  This is the famous old cosmological constant problem, for a review see \cite{Weinberg:1988cp, Sahni:1999gb, Peebles:2002gy, Copeland:2006wr}.  In addition, there is a new cosmological constant problem stating why the effects of dark energy  become relevant at this very late stage of the expansion history of the Universe, at redshift around $z \sim 0.3$. 

While imposing a cutoff by hand to regularize the vacuum zero point energy is a useful approach to start with, but it is not technically correct. The simple reason is that it violates the underlying local Lorentz invariance when a cutoff scale in momentum space is introduced. Therefore, to regularized quantum infinities in the presence of GR, one has to employ a prescription which keeps the general covariance intact. There are a number of well-established schemes for regularization and renormalization infinities in curved spacetimes which respect this fundamental requirement. 
These includes  the point splitting regularization method  \cite{Christensen:1976vb, Christensen:1977jc, Davies:1977ze, Anderson:1990jh} and the dimensional regularization (DR) procedure \cite{tHooft:1972tcz, tHooft:1974toh,
Bollini:1972ui,  Deser:1974cz, Dowker:1975tf, Barvinsky:1985an, Onemli:2002hr, Brunier:2004sb, Miao:2005am, Prokopec:2008gw, Miao:2010vs, Glavan:2021adm, Glavan:2020gal}. The DR scheme is particularly useful in GR as it automatically can be implemented within general covariance of GR. In this prescription, one considers a curved $D$-dimensional spacetime and studies the quantum perturbations. As in flat background, one encounters infinities at the coincident limits, corresponding to the inherited  UV divergence in quantum mechanics. To regularize the UV divergence in physical quantities such as $\langle \rho_v \rangle$, one considers $D=4-\epsilon$ and looks at the possible singular terms which appear with inverse powers of $\epsilon$. 
The regularization corresponds to removing these divergent terms with inverse powers of $\epsilon$ while a proper renormalization requires the inclusion of counter terms in the starting Lagrangian in order to absorb the singular terms from physical quantities. Following this logic, there is no restrictions in employing DR to quantum fields in a classical curved backgrounds as long as the starting Lagrangian and the counter terms respect the underlying 
general covariance. This was employed systematically in the literature of quantum field theory in curved spacetime such as   in works listed above and many other works.

Employing DR scheme in flat background for a quantum field  with mass $m$, one actually obtains \cite{Martin:2012bt, Akhmedov:2002ts, Ossola:2003ku}, 
\ba
\label{rho-flat}
\langle \rho_v\rangle = \frac{m^4}{64 \pi^2} \ln\big(\frac{m^2}{\mu^2} \big)\, ,
\ea
in which $\mu$ is a regularization mass scale. Note that $\mu$ can be a physical scale such as the mass of a given field like electron or the scale of Hubble expansion rate at a given epoch etc. However, it is fixed once 
and with the same value for all fields in the spectrum. In the presence of multiple fields with masses $m_i$ such as in SM spectrum, the total contribution in $\langle \rho_v\rangle$ is the sum of  Eq. (\ref{rho-flat}) from each field with appropriate $\pm1$ signs for the bosons and fermions. This shows that the contribution of a fundamental field with mass $m$ in vacuum energy is at the order $m^4$. The heavier is the field, the higher is its contribution in vacuum energy density. 

It is natural to look for vacuum zero point energy in a curved background. However, in a curved spacetime  the solutions for the mode functions are non-trivial. In addition, the notion of vacuum is non-trivial in a curved  background \cite{Hawking:1975vcx, Unruh:1976db, Unruh:1983ms, Jacobson:2003vx, Firouzjahi:2022rtn}.   The vacuum zero point energy and its regularizations in a dS  background are vastly studied using either DR  or other regularization schemes. For an incomplete list of papers see for example  the works of Woodard and collaborators 
\cite{ Onemli:2002hr, Onemli:2004mb, Miao:2005am, Janssen:2008px, Glavan:2021adm}  who have mainly employed the DR scheme for regularization and renormalization.  Among many things, it is shown that in a dS background with the Hubble expansion rate $H$, the contribution of the massless and light fields in vacuum energy is at the order $H^4$. This is in contrast to the flat background where from 
Eq. (\ref{rho-flat}) one concludes that the massless fields do not contribute into vacuum energy density. However, for very heavy fields with $m \gg H$, it is shown in \cite{Firouzjahi:2022xxb, Firouzjahi:2023wbe, Firouzjahi:2022vij} that $\langle \rho_v\rangle$ obeys the same formula as Eq. (\ref{rho-flat}) with subleading $O(m^2 H^2)$ corrections.  This may be understood from local Lorentz invariance and equivalence principle. 
 
Another question of interest is to look at the fluctuations of the vacuum zero point energy itself, $\delta \rho_v$. This was studied in more details in \cite{Firouzjahi:2022xxb, Firouzjahi:2023wbe, Firouzjahi:2022vij} where it is shown that the fluctuations in the vacuum zero point energy is large in the sense that $\delta \rho_v \sim \langle \rho_v \rangle$. Furthermore, it is shown that the distribution of the vacuum zero point energy is highly non-Gaussian in which $\delta \rho_v^3 \sim \langle \rho_v \rangle^3$. In this work we study the effects  of the vacuum zero point energy and its fluctuations in an inflationary background. By considering the perturbations of the vacuum zero point energy and requiring that the primordial perturbations to be   nearly Gaussian and under perturbative control, we obtain a  strong upper bound on the mass of the fundamental fields during inflation. While in this work we investigate the effects of vacuum zero point fluctuations to put constraints on the mass of fundamental fields, but the question of investigating the masses and couplings of fundamental fields during inflation were  investigated extensively in the context of cosmological collider physics, for an incomplete list of papers on this direction see \cite{Arkani-Hamed:2015bza, Meerburg:2016zdz, Chen:2016uwp, Wang:2019gbi, Chen:2009zp, Noumi:2012vr, Emami:2013lma}. \\


\section{Quantum Fields in Inflationary Background}
\label{QFT-inflation}

In this section we review the quantum field perturbations in inflationary background. This analysis follow the earlier works \cite{Firouzjahi:2022xxb, Firouzjahi:2023wbe}. 

We consider a  scalar field $\chi$ with mass $m$ which is minimally coupled to gravity. The background is an inflationary universe which is driven by the inflaton field $\phi$.  While the inflaton field rolls slowly along its classical potential $V(\phi)$, the field $\chi$ is stuck in its local minimum with no classical evolution. However, its is under quantum fluctuations which contribute to its vacuum energy density. We assume that the vacuum zero point energy associated to the  spectator field does not dominate the background inflation dynamics.  This sets an upper bound on  the mass of $\chi$ field. As usual, we assume that the total cosmological constant from inflaton and the spectator  field is set to zero at the end of inflation. This is another realization of the old cosmological constant problem
where one requires the potential to be zero or small for a consistent expansion history of the Universe during the hot big bang cosmology. 
While we perform the analysis for a single fundamental scalar field, but our results  can be extended to other fundamental fields with various spins.  

As discussed before, in order to regularize the UV divergences associated to vacuum zero point energy, we employ the DR scheme and consider a $D$-dimensional inflationary background. To simplify further, we assume the background is  nearly a dS spacetime as in standard slow-roll inflationary setups. 

The background metric is  a $D$-dimensional FLRW universe with the line element,  
\ba
ds^2 = a(\tau)^2 \big( -d \tau^2 + d {\bf x}^2 \big) \, ,
\ea
where  $a(\tau)$ is the  scale factor and $\tau$ is the conformal time which is related to the cosmic time via $d \tau = dt/a(t)$. In our approximation of a near dS background, we have $a H \tau =-1 $ in which $H$ is the Hubble expansion rate during inflation which is constant in our approximation. In the above metric, $d  {\bf x}^2$ represents the line element along the $D-1$ spatial dimensions.  

To study  the quantum perturbations, we  introduce the canonically normalized  field $\sigma(x^\mu)$
\ba
\mathrm{\sigma(x^\mu)}\equiv a^{\frac{D-2}{2}}\chi(x^\mu) \, ,
\ea
and expand its quantum perturbations   in the Fourier space as follows,
\begin{equation}
\label{quantized-sigma}
\sigma\left(x^\mu\right)=\int \frac{d^{D-1} \mathbf{k}}{(2 \pi)^{\frac{(D-1)}{ 2}}}\left(\sigma_k(\tau) e^{i \mathbf{k}\cdot\mathbf{x}} a_{\mathbf{k}}+\sigma^*_k(\tau) e^{-i \mathbf{k} \cdot \mathbf{x}} a_{\mathbf{k}}^{\dagger}\right)\,,
\end{equation}
in which $\sigma_k(\tau)$ is the quantum mode  function while
$a_{\mathbf{k}}$ and $a_{\mathbf{k}}^{\dagger}$ are the annihilation and creation operators  satisfying the following
commutation relation in $D-1$ spatial dimension,
\begin{equation}
\label{Noncommutative}
\left[a_{\mathbf{k}}, a_{\mathbf{k^\prime}}^{\dagger}\right]=\delta^{D-1}(\mathbf{k}-\mathbf{k^\prime})\,.
\end{equation}

In terms of the canonically normalized field $\sigma$, the Klein-Gordon field equation takes the following form,
\begin{equation}
\label{EquationofMotion}
\sigma_k^{\prime \prime}(\tau)+\left[k^2+ \frac{1}{\tau^2} \Big(
\frac{m^2}{H^2 } -\frac{D(D-2)}{4} \Big) \right] \sigma_k(\tau)=0\,.
\end{equation}
The above equation is similar to the Mukhanov-Sasaki equation in 
$D$-dimension dS background. 

Imposing the Bunch-Davies (Minkowski) vacuum deep inside the horizon, the
solution for the mode function  is obtained in terms of the Hankel function
\begin{equation}
\label{chi-k}
\chi _k(\tau) = a^{\frac{{2 - D}}{2}}{\sigma _k}(\tau ) = {( - H\tau )^{\frac{{D - 1}}{2}}}{\left( {\frac{\pi }{{4H}}} \right)^{\frac{1}{2}}} {{e^{\frac{i \pi}{2}  (\nu + \frac{1}{2})} }}
H_\nu ^{(1)}( - k\tau ){\mkern 1mu}\,,
\end{equation}
where
\begin{equation}
\label{nu-00}
{{
\nu  \equiv  \frac{1}{2}{\mkern 1mu} \sqrt {(D-1)^2- 4 \beta^2}\, , \quad \quad
\beta \equiv \frac{m}{H} }}\, .
\end{equation}
From the above expression  we see that $\nu$ can be either real or pure imaginary, depending on the mass $m$. For a light field with $\beta <1$,    $\nu$ is real while for a heavy field with $\beta \gg 1$ it is a pure complex number. \\


\subsection{Vacuum Zero Point Energy}
\label{vac-reg}

We are interested in the vacuum zero point energy $\rho_v$ associated to $ \chi$ quantum fluctuations. It is convenient to define the following 
components of $\rho_v$, 
\ba
\label{rhoi}
\rho_1 \equiv \frac{1}{2} \dot \chi^2 \, , \quad \quad
\rho_2 \equiv \frac{1}{2} g^{i j} \nabla_i \chi \nabla_j \chi \, , \quad \quad
\rho_3 \equiv \frac{1}{2} H^2 \chi^2 \, ,
\ea
so
\ba 
\label{rho-tot}
\rho_v= \rho_1+ \rho_2 +\beta^2 \rho_3 \, .
\ea 
Note our convention in which we have pulled out a factor $\beta=m/H$ when defining $\rho_3$ so $\beta^2 \rho_3= \frac{1}{2} m^2 \chi^2$. 

We would like to calculate the vacuum expectation values like 
$\langle \rho_v \rangle \equiv \langle 0| \rho_v |0\rangle$ in which $|0\rangle$
is the vacuum of the free theory i.e. the Bunch-Davies vacuum. Here we briefly outline the analysis, for more details see \cite{Firouzjahi:2023wbe}.  

Let us start with $\langle \rho_1\rangle $. With the mode function given in Eq. (\ref{quantized-sigma}) we obtain
\ba
\langle \rho_1\rangle  =\frac{ \mu ^{4 - D} }{2a^2(\tau ) }
\int \frac{ d^{D - 1} {\bf k}}{(2\pi )^{D - 1} }  \left| \chi _k^\prime (\tau ) \right|^2 \, ,
\ea
in which $\mu$, as in standard DR analysis, is a mass scale to keep track of the dimensionality of the physical quantities. 

To calculate the integral, we decompose it into the radial and angular parts as follows
\ba
{{\rm{d}}^{D-1}}{\bfk} = {k^{D - 2}}\;{\rm{d}}k\; {{{\rm{d}^{D-2}} }}\Omega {\mkern 1mu}\, ,
\ea
in which  ${\rm{d}^{D-2}}\Omega$ represents the volume of the 
$D-2$-dimensional  angular part, 
\ba
 \int \mathrm{d}^{D-2} \Omega=\frac{ 2 \,  \pi^{\frac{D-1}{2}}}{\Gamma\left(\frac{D-1}{2}\right)}\,.
\ea
Defining the dimensionless variable $x\equiv - k \tau$ and  combining all numerical factors,  we end up with the following integral,
\ba
\label{rho1}
\langle \rho_1 \rangle =
{\frac {{\pi}^{\frac{3-D}{2}  }{\mu}^{4-D}{H}^{D}}{{2}^{1+D}\Gamma \left( \frac{D-1}{2} \right) }} e^{-\pi \mathrm{Im}(\nu)}  \int_0^{\infty} dx~x  \left|\frac{d}{d x}\left(x^\frac{D-1}{2} H_{ \nu}^{(1)}(x)\right)\right|^2 \, .
\ea
Performing the integral \footnote{We use the Maple computational software to calculate the integrals like in Eq. (\ref{rho1})}, we obtain 
\begin{eqnarray}\label{rho1-int}
\left\langle {{\rho _1}} \right\rangle  = \frac{{{\mu ^{4 - D}}{\pi ^{ - \frac{D}{2} - 1}}}}{4} {\mkern 1mu} \Gamma \Big( {\nu  + \frac{D}{2} + \frac{1}{2}} \Big)\Gamma \Big( { - \nu  + \frac{D}{2} + \frac{1}{2}} \Big)\Gamma \big( { - \frac{D}{2}} \big) \cos \big( {\pi {\mkern 1mu} \nu } \big)   {\big( {\frac{H}{2}} \big)^D}\,.
\end{eqnarray}

Performing the same steps  for $\langle \rho_2 \rangle$ and 
$\langle \rho_3 \rangle$, one can show that the following relations hold,
\ba
\label{rho13}
\langle \rho_1 \rangle =  \frac{\beta^2}{D}  \langle \rho_3 \rangle \, , \quad \quad \langle \rho_2 \rangle = -(D-1) \langle \rho_1 \rangle =
- \frac{(D-1)}{D}  \beta^2  \langle \rho_3 \rangle  \, .
\ea
The above relations between $\langle \rho_i \rangle$ will be useful later on. 

Plugging the above expressions for $\langle \rho_i \rangle$ in $\langle \rho_v \rangle$ in Eq. (\ref{rho-tot}), we obtain
\ba
\label{rho-3}
\langle \rho_v \rangle = \frac{2 \beta^2}{D}  \langle \rho_3 \rangle \, .
\ea
Following the same steps, one can check that the following relation between the pressure $P$ and the energy density holds \cite{Firouzjahi:2023wbe}, 
\ba
\label{P-rho}
\langle P_v \rangle =- \langle \rho_v \rangle \, .
\ea
This is an important result. It shows that the vacuum zero energy has the form of a cosmological constant. This is physically consistent since we calculate the contribution from the  bubble Feynman diagrams and  Lorentz invariance is expected to hold locally  with $\langle T_{\mu \nu} \rangle = \langle \rho_v \rangle g_{\mu \nu}$. 

The above result for  $ \langle \rho_v \rangle$ is valid for a general value of $D$. Now, we perform the DR by setting $D= 4-\epsilon$ and expand $ \langle \rho_v \rangle$ to leading orders in powers of $\epsilon$. As usual, the UV divergent contributions are controlled by the singular pole term $\epsilon^{-1}$ which should be absorbed by appropriate counter terms. Regularizing this divergence contribution, the remaining finite  contribution is interpreted as the renormalized energy density,  obtained to be \cite{Firouzjahi:2023wbe}
\ba
\label{rho-reg}
\langle \rho_v \rangle_{\mathrm{ren}}  
= \frac{H^4 \beta^2}{64 \pi^2} \Big \{ ( \beta^2  -2)
\Big[ \ln\Big( \frac{H^2}{4\pi \mu^2 }  \Big) + 2\Psi(\nu+ \frac{1}{2})  -  \pi \tan( \nu \pi)  \Big] + 1 - \frac{3}{2} \beta^2  \Big \} \, ,
\ea
in which $\Psi(x)$ is the digamma function and $\nu$ is now given by setting $D=4$ in Eq. (\ref{nu-00}),
\ba
\label{nu-val}
\nu = \frac{1}{2}\sqrt{9- 4 \beta^2 } \, .
\ea

The appearance of $\ln\Big( \frac{H}{ \mu }  \Big)$ in $\langle \rho_v \rangle_{\mathrm{reg}}$ is the hallmark of quantum corrections from DR scheme.  To read off the physical contribution, we need to impose the conditions of renormalization such that the physical quantities do not depend
on the parameter $\mu$.  This can be achieved by choosing a physical value for the mass scale parameter $\mu$ or if we compare the values of $\langle \rho \rangle_{\mathrm{reg}}$ at two different energy scales and examine its running with the change of the energy scale. We impose our renormalization condition more specifically later on. 

As mentioned previously, depending on the mass of the field, $\nu$ in Eq. (\ref{nu-val}) can be either real or imaginary. For light enough mass with 
$\beta \leq \frac{3}{2}$ it is real while for heavier field it is pure complex.  

Let us look at the value of $\langle \rho \rangle_{\mathrm{reg}}$ in Eq. (\ref{rho-reg}) for some limiting cases. For a massless field with $\beta=0$, we obtain 
\ba
\label{massless-rho}
\langle \rho_v \rangle_{\mathrm{ren}}= \frac{3 H^4}{32 \pi^2} \, ,
 \quad \quad \quad  (\beta=0) \, .
\ea
This shows that for the massless field, the vacuum  energy density scales like $H^4$. On the other hand, for very heavy field with $\beta \gg 1$, one obtains \cite{Firouzjahi:2022xxb, Firouzjahi:2023wbe}
\ba
\label{rho-reg-heavy2}
\langle \rho_v \rangle_{\mathrm{ren}}
= \frac{m^4}{64 \pi^2}  \ln\big( \frac{m^2}{4\pi \mu^2 }  \big)
+{ \cal O} ( m^2 H^2) \,  \quad \quad (\beta \gg 1) \, .
\ea
This has exactly the same form as in flat background Eq. (\ref{rho-flat}). As explained before, this may be expected from local Lorentz invariance and the equivalence principle.  Finally, for the intermediate mass range with $\beta \lesssim 1$, the vacuum energy has the form (\ref{rho-reg-heavy2}) but with a numerical prefactor depending on  $\beta$ as well. 

As we would like to put an upper bound on the mass of the field during inflation, we consider the heavy and intermediate mass range fields.  For light field where  $\langle \rho_v \rangle \sim H^4 $, the contribution of the vacuum energy in  inflation dynamics is negligible as $H^4 \ll 3 M_P^2 H^2$. \\


\subsection{Fluctuations in Vacuum Zero Point Energy }
\label{vac-fluc}
As observed in \cite{Firouzjahi:2022xxb, Firouzjahi:2023wbe, Firouzjahi:2022vij}, the vacuum energy is subject to random fluctuations with large amplitudes. 
 Denoting the statistical variation of 
$\rho_v$ by $\delta \rho_v^2 \equiv \langle \rho_v^2 \rangle - \langle \rho_v \rangle^2$, it is shown  in \cite{Firouzjahi:2022xxb, Firouzjahi:2023wbe, Firouzjahi:2022vij} that 
\ba
\label{rho-contrast2}
{\delta \rho_v^2} =10  {\langle \rho_v \rangle^2}\, .
\ea
The above result for the  density contrast of the vacuum energy holds in both flat,  dS  as well as  in  black hole backgrounds. This result plays crucial role in our investigation of the upper bound on the mass of the quantum fields during inflation.  The fact that $\delta \rho_v \sim
\langle \rho_v \rangle$ indicates that the distribution of the vacuum zero pint energy is non-linear and non-perturbative, which may generate  inhomogeneity  and anisotropies on small scales, see also  \cite{Wang:2017oiy, Cree:2018mcx, Wang:2019mbh, Wang:2019mee, Wang:2023tzm}. 

Since  $\rho^2$ is a composite operator, one may argue that its regularization may be independent of the regularization associated to the operator $\rho$.
However, the key point is that both of these operators are based on the Gaussian field $\chi$. For example, consider $\rho_3 \sim \chi^2$. Then,
$\rho_3^2 \sim \chi^4$. Therefore, the regularization of 
$\langle \rho_3^2\rangle$ will not be independent of the regularization of 
$\langle \rho_3\rangle$. Indeed, since both of these operators are related to the same Gaussian field $\chi$, then it is natural to expect that 
$\langle \rho_3^2\rangle \sim \langle \rho_3\rangle^2$.

In addition, it is shown in \cite{Firouzjahi:2023wbe} that the fluctuations in the distribution of the vacuum zero point energy is highly non-Gaussian. Denoting the  skewness  in vacuum zero point energy by $\delta \rho_v^3 \equiv \big\langle \big( \rho_v  - \langle \rho_v \big\rangle\big)^3\rangle$, one obtains  that\footnote{In  \cite{Firouzjahi:2023wbe} a slightly different definition of skewness is used, defined by $\delta \rho_v^3 \equiv \langle \rho_v^3\rangle  - \langle \rho_v \rangle^3$.}
\ba
\label{non-G}
{\delta \rho_v^3}  = 62  {\langle \rho_v \rangle ^3}\, .
\ea

Eq. (\ref{non-G}) indicates that the distribution of the fluctuations of the vacuum zero point energy is highly non-Gaussian. This will play important roles in our analysis of obtaining an upper bound for the mass of the fundamental fields.

Before closing this section, we comment that while we have obtained the density contrast and the skewness associated to zero point fluctuations of an spectator field, but the same results apply for inflaton field as well. The reason is that all we needed was the solution of the mode function (\ref{chi-k}) which works for all fields, whether light or heavy. The case of inflaton corresponds to $\beta \ll1$. Correspondingly,  the density contrast and the skewness for the zero point fluctuations of inflaton satisfy the same expressions as Eqs. (\ref{rho-contrast2}) and (\ref{non-G}). \\


\subsection{ Renormalization Condition} 
\label{renormalization}

As discussed above, to read off the physical quantities we should impose the renormalization condition so the final results do not depend on the parameter 
$\mu$. Our renormalization condition is that the total vacuum zero point energy induced from all quantum fields to vanish during and at the end of inflation. This requirement guarantees that inflation is driven by the classical inflaton potential and is terminated when the usual slow-roll conditions are violated.  Furthermore, the Universe undergoes the standard hot big bang phase after (p)reheating. Of course, one may tune the final vacuum energy density to a very small value as required today for the source of dark energy. But this requires extreme fine-tuning which is nothing but the old cosmological constant problem. 

It is important that our renormalization condition is imposed on the total vacuum energy density and not for an individual field. More specifically, suppose we have $N$ scalar fields during inflation in which one of them is the inflaton field itself, denoted by $\phi$, while the remaining fields are collectively denoted by $\chi_a$ with $a=1,..., N-1$. We denote all fields by 
$\{ \chi_i\}= \{\phi, \chi_a \}$. 

Neglecting the subleading contribution, the total vacuum zero point energy from all massive fields $\{ \chi_i\}$   from  Eq. (\ref{rho-reg-heavy2}) is given by 
\ba
\label{rho-reg-heavy3}
\langle \rho_v \rangle_{\mathrm{tot}}
= \sum_i^{N}\frac{m_i^4}{64 \pi^2}  
\ln\big( \frac{m_i^2}{4\pi \mu^2 }  \big) \, ,
\ea
in which $m_i$ is the mass of $ \chi_i$. It is important that the parameter 
$\mu$ is the same for all field as we impose the DR on the total energy density.

Our renormalization condition is that the total vacuum zero point energy to vanish, $\langle \rho_v \rangle_{\mathrm{tot}}=0$. This fixes $\mu$ in terms of 
$m_i$ as follows,
\ba
\label{mu-val}
 \ln \mu^2 = \frac{\sum_i^{N} m_i^4 \ln \big(\frac{m_i^2}{4 \pi}\big)}{\sum_i^N m_i^4} \, .
\ea
Now, using the above value of $\mu$ for the vacuum zero point energy associated to each field,  Eq. (\ref{rho-reg-heavy2}) yields
\ba
\label{rho-i}
\langle \rho_i\rangle_{\mathrm{ren}}= \frac{m_i^4}{64 \pi^2 \sum_k m_k^4}  
\sum_{j=1}^N  m_j^4\, \ln \big(\frac{m_i^2}{m_j^2}\big) \, .
\ea

Employed with the above renormalization condition, we can also calculate the fluctuations in the distributions of the total energy density which for simplicity we denote by $\delta \rho_v$.  
Note that the density contrast and skewness in  Eqs. (\ref{rho-contrast2}) and (\ref{non-G}) are for a single field. But we are interested in the fluctuations in total energy density.
As the fields $\chi_i$ are statistically independent and we assume there is no direct interactions between them, then the perturbations in total energy density is given by
\ba
\label{delta-rho-tot}
\delta \rho_v^2 &=& \sum_i \big( \delta \rho_{v}^{(i)}\big)^2 = 10 \sum_i \langle \rho_i \rangle^2  \nonumber\\
&=&  10 \sum_{i, j=1}^{N} \Big[  \frac{m_i^4 m_j^4}{64 \pi^2 \sum_k m_k^4}  
 \ln \big(\frac{m_i^2}{m_j^2}\big) \Big]^2 \, .
\ea
Similarly, the skewness in total energy density is given by,
\ba
\label{skew-rho-tot}
\delta \rho_v^3 &=& \sum_{i=1}^{N} \big( \delta \rho_{v}^{(i)}\big)^3 = 62 \sum_i \langle \rho_i \rangle^3  \nonumber\\
&=&  62 \sum_{i, j=1}^{N} \Big[  \frac{m_i^4 m_j^4}{64 \pi^2 \sum_k m_k^4}  
 \ln \big(\frac{m_i^2}{m_j^2}\big) \Big]^3 \, .
\ea 

To have a feeling of the above results, let us consider two specific cases in detail. First, suppose we have only two fields $N=2$, the inflaton field $\phi$ and the heavy field $\chi$. In this case,
\ba
\langle \rho_v^{(\chi)} \rangle = -  \langle \rho_v^{(\phi)} \rangle 
= \frac{m_\chi^4 m_\phi^4}{64 \pi^2 ( m_\chi^4 + m_\phi^4)} 
\ln \big( \frac{m_\chi^2}{m_\phi^2} \big)  \,  ,  \quad \quad (N=2) .
\ea
Now suppose that the field $\chi$ is significantly heavier than the inflaton field 
which is the limit we would be interested. In this case, the above expression can be approximated to,
\ba
\label{light-cont}
  \langle \rho_v^{(\chi)} \rangle \simeq 
  \frac{ m_\phi^4}{64 \pi^2 } 
\ln \big( \frac{m_\chi^2}{m_\phi^2} \big) \, \,  ,  \quad \quad (N=2) .
\ea
Interestingly, we see that the scale of the vacuum energy density of the heavy field is actually dictated by the mass of the light (inflaton) field. This is a direct implication of the renormalization condition that 
$\langle \rho_v \rangle_{\mathrm{tot}}=0$. 

The situation becomes more non-trivial if more than one heavy fields are present in the spectrum. For example, suppose we have three fields, the inflaton $\phi$ and two heavy spectator fields $\chi_1$ and $\chi_2$. The vacuum energy density of the inflaton and either of the heavy fields, say $\chi_1$, 
are given by,
\ba
 &&\langle \rho_v^{(\phi)} \rangle  =
 \frac{m_\phi^4 }{64 \pi^2 (  m_\phi^4+ m_{\chi_1}^4+ m_{\chi_2}^4 )} 
\Big[m_{\chi_1}^4 \ln \big( \frac{m_\phi^2}{m_{\chi_1}^2} \big)  + m_{\chi_2}^4 \ln \big( \frac{m_\phi^2}{m_{\chi_2}^2} \big) 
\Big] \, , \\
\label{multiple-chi}
&&\langle \rho_v^{(\chi_1)} \rangle  =
 \frac{m_{\chi_1}^4 }{64 \pi^2 (  m_\phi^4+ m_{\chi_1}^4+ m_{\chi_2}^4 )} 
\Big[m_\phi^4 \ln \big( \frac{m_{\chi_1}^2}{m_\phi^2} \big)  + m_{\chi_2}^4 \ln \big( \frac{m_{\chi_1}^2}{m_{\chi_2}^2} \big)  \Big] \, .
\ea
Now suppose we have $m_\phi \ll m_{\chi_1},  m_{\chi_2}$. In this case, neglecting the logarithmic contributions, we see that $\langle \rho_v^{(\phi)} \rangle \propto m_\phi^4$ which may be expected. However, for the heavy fields we obtain  $\langle \rho_v^{(\chi_a)} \rangle \propto \frac{m_{\chi_a}^4 m_\phi^4}{(m_{\chi_1}^4+ m_{\chi_2}^4 )}$. In the case $m_{\chi_1} \sim m_{\chi_2}$, this yields the result that 
$\langle \rho_v^{(\chi_a)} \rangle \propto m_{\chi_a}^4 $. However, if we consider the case where $m_\phi\ll m_{\chi_1} \ll m_{\chi_2}$, then 
$\langle \rho_v^{(\phi)} \rangle \propto m_\phi^4$ and $  \langle \rho_v^{(\chi_1)} \rangle \simeq - \langle \rho_v^{(\chi_2)} \rangle  \propto m_{\chi_1}^4$.  The conclusion is that when there is a hierarchy in the masses of the fields, then the lighter fields at each step in the hierarchy will fix the scale of the vacuum zero point energy densities associated to the fields at the corresponding step in the hierarchy. \\

\section{Upper Bounds on the Masses and Couplings}
\label{bound}

The quantum fluctuations of inflaton and other fields are the source of observed structures in Universe. These perturbations are created on all scales. On long IR scales they can be directly linked to the observed large scale structure of cosmos but on small (UV) scales their roles are more subtle. At the background level the UV scale perturbations  lead to the cosmological constant problem. However, as  discussed in previous section, these perturbations also induce fluctuations in distributions of the vacuum zero point energy itself.  Correspondingly,  the fluctuations in vacuum zero point energy contribute to primordial curvature perturbations and can affect the cosmological observables such as the power spectrum and bispectrum.

With the above discussions in mind, we are ready to put constraints on the mass of the quantum fields during inflation. To be specific, here we consider the case where we have $N\ge 2$ non-interacting scalar fields in which one of them, the inflaton field $\phi$, is light while the remaining spectator fields $\chi_a$ are relatively heavy with $m_{\chi_a} >H$.   This setup can be extended to more general cases involving multiple spectator fields with different spins and hierarchy of masses.   Our implicit assumption is that the spectator field $\chi_a$ are heavy enough compared to $H$.  As they they are locked in their local minima with zero 
potential, they have no classical values and with no classical  contributions  in energy density.

It is assumed that the inflationary background is driven with the inflaton field  with a classical potential $V(\phi)$.  All fields are subject to quantum fluctuations so we should take into account  their contributions in vacuum zero point energy, i.e. $\rho_v = \rho_v^{(\phi)} + \sum_a\rho_v^{(\chi_a)}$ with $a=1,..., N-1$.   The total energy density driving the background expansion  is the sum of the inflaton classical energy density $\rho_\phi^{\mathrm{cl}}$
and the  vacuum zero point energy of both fields, i.e. 
$ \rho_{\mathrm{tot}}= \rho_\phi^{\mathrm{cl}}+ \langle \rho_v \rangle$. In the slow-roll approximation we 
further have $\rho_\phi^{\mathrm{cl}} \simeq V(\phi)$. \\

\subsection{Bounds from Power Spectrum}

From our studies of perturbations in vacuum zero point energy, we can obtain upper bounds on the mass of heavy spectator fields  during inflation.  In our view, the vacuum energy density from the quantum fields  provides random fluctuations in energy density.  Observationally, the perturbations in energy density is the source of perturbations in CMB map with the amplitude 
$\frac{\delta \rho}{\rho} \sim \frac{\delta T}{T }\sim 10^{-5}$. In the analysis below, we look at the perturbations in real space. It is understood that the perturbations from vacuum zero point energy have the correlation length $m^{-1}$ so if the field is heavy, its correlation length  is sub-Hubble. However, the key issue is that we look at the accumulative contributions of all UV modes when looking at statistical quantities such as $\langle \big(\frac{\delta \rho({\bf x})}{\rho}\big)^2 \rangle$ locally in real space. In the spirit, this is similar to the cosmological constant problem at the background level when all UV modes contribute to the averaged quantity $\langle \rho_v \rangle$. Now, we extend this view to perturbation in vacuum energy density itself. 

To start, let us consider the curvature perturbation on 
surface of constant energy density $\zeta$, defined as \cite{Lyth:2004gb, Bassett:2005xm}
\ba
\zeta \equiv -\psi + \frac{H}{\dot \rho_{ \mathrm{tot}} } 
\Delta \rho \, ,
\ea
in which $\psi$ is the curvature perturbation on three-dimensional spatial hypersurface.  Also,   $\rho_{ \mathrm{tot}} $ is the total background energy density while $\Delta \rho$  is the perturbation in total energy density. Note that neither $\Delta \rho$ nor $\psi$ are gauge invariant but 
$\zeta $ is. To simplify further, we go to spatially flat  gauge where $\psi=0$. 

As mentioned previously, in our case of interest $\rho_{ \mathrm{tot}} = \rho^{\mathrm{cl}}_\phi + \langle \rho_v\rangle$. Since  $ \langle \rho_v\rangle$ is the vacuum energy density which is constant by construction,  we conclude that 
\ba
\label{rho-dot}
\dot \rho_{ \mathrm{tot}} = \dot \rho^{\mathrm{cl}}_\phi  = - 3 H (\rho^{\mathrm{cl}}_\phi + P^{\mathrm{cl}}_\phi ) 
=-3 H \dot \phi^2 = -6 \epsilon_H M_P^2 H^3 \, ,
\ea
in which $\epsilon_H \equiv -\dot H/H^2$ is the first slow-roll parameter. 

On the other hand, the fluctuations in energy density receive contributions from both $\phi$ and $\chi_a$ fields, $\Delta \rho= \Delta \rho_\phi + \Delta \rho_\chi$, in which we have defined $\Delta \rho_\chi = \sum_a \Delta \rho_{\chi_a}$. Correspondingly, the total curvature perturbation on flat slicing $\psi=0$ is given by, 
\ba
\label{zeta-flat}
\zeta_{\mathrm{flat}}=  \frac{H}{ \dot \rho^{\mathrm{cl}}_\phi} \big( \Delta \rho_\phi + \Delta \rho_\chi \big) \, .
\ea
As mentioned before,  the spectator fields have no classical components and  their  contributions are from the vacuum zero point fluctuations, 
\ba
\label{delta-rho-chi}
\Delta \rho_\chi = \sum_a \rho_v^{(\chi_a)} - \langle \rho_v^{(\chi_a)} \rangle 
\equiv \Delta \rho_v^{(\chi)} \, .
\ea
In terms of $ \chi_a$ quantum fluctuations, $\Delta \rho_\chi$ is second order in $ \chi_a^2$ or its derivatives like $\dot \chi_a^2$.  This is because $ \chi_a$ is pure quantum perturbations with $\langle \chi_a \rangle=0$. Also note that 
$\langle \Delta \rho_v^{(\chi)}   \rangle =0$. 

On the other hand,  $\Delta \rho_\phi $ can have a linear contribution in 
$\delta \phi$. This is because $\phi$ is rolling  on its classical potential so
$\Delta \rho_\phi $ can have mixed contributions such as $\dot \phi  \delta \dot \phi$ or $m_\phi^2 \phi \delta \phi$ etc. We denote this mixed contribution which is linear in $\delta \phi$ perturbations by $\Delta \rho_\phi^{(1)}$.  This is the standard source of perturbation in inflationary energy density in the absence of vacuum zero point energy. Similar to Eq. (\ref{delta-rho-chi}), the  contribution of vacuum zero point energy in $\Delta \rho_\phi $,  which is second order in $\delta \phi$, is denoted by $\Delta \rho_v^{(\phi)}$, 
\ba
\label{delta-rho-phi}
\Delta \rho_v^{(\phi)} \equiv \rho_v^{(\phi)} - \langle \rho_v^{(\phi)} \rangle \, .
\ea
Combining the standard contribution $ \Delta \rho_\phi^{(1)}$ and the contributions from the vacuum zero point fluctuations of $\Delta \rho$ in Eq. (\ref{zeta-flat}), and discarding the subscript ``flat" for convenience, we obtain
\ba
\label{zeta-flat2}
\zeta=  \frac{H}{ \dot \rho^{\mathrm{cl}}_\phi} 
\big( \Delta \rho_\phi^{(1)} + \Delta \rho_v^{(\phi)} + 
\Delta \rho_v^{(\chi)}\big) \, .
\ea

The two point correlation functions  $\langle \zeta^2 \rangle$ gives the amplitude of temperature fluctuations in CMB maps.  The cosmological observations such as the Planck observation indicate that the curvature power spectrum  is nearly scale invariant and Gaussian  \cite{Planck:2018jri}.  This is because the inflaton potential is nearly flat (i.e. the background is nearly dS) and the inflaton is light compared to Hubble expansion rate, $m_\phi \ll H$. The first term in Eq. (\ref{zeta-flat2}) yields the usual nearly scale-invariant power spectrum. However, the remaining two terms, 
$\Delta \rho_v^{(\phi)} $ and $ \Delta \rho_v^{(\chi)}$, originating from the vacuum zero point fluctuations, have non-trivial scale-dependence in Fourier space. This is because they are in the form $\delta \phi^2$ and $\delta \chi^2$. In addition, the index $\nu$ for the spectator field is far from the special value $\nu=\frac{3}{2}$ ($\nu$ can even become complex-valued) 
 if $\chi_a$ is heavy. Therefore, their contributions will modify the near scale-invariance of the standard power spectrum coming from the first term  in Eq. (\ref{zeta-power}). We present the analysis of the  scale-dependence of 
 the contributions of $\Delta \rho_v^{(\phi)} $ and $ \Delta \rho_v^{(\chi)}$ in power spectrum in Appendix \ref{appendix}.

Using the above expression for $\zeta$, and noting that $\langle  \Delta \rho_\phi^{(1)} \rangle=0$  (since it is linear in terms of $\delta \phi$ fluctuations), we obtain
\ba
\langle \zeta^2 \rangle = \big( \frac{H}{ \dot \rho^{\mathrm{cl}}_\phi}\big)^2
\Big[ \big \langle   (\Delta \rho_\phi^{(1)})^2  \big \rangle  + \big \langle   (\Delta \rho_v^{(\phi)})^2  \big \rangle + \big \langle   (\Delta \rho_v^{(\chi)})^2  \big\rangle   + 2  \big \langle  \Delta \rho_\phi^{(1)} \Delta \rho_v^{(\phi)}  
\big \rangle + 2 \big \langle  \Delta \rho_v^{(\chi)} \Delta \rho_v^{(\phi) }  \big \rangle   \Big]\, .
\ea 
As the $\delta \phi$ and $\chi_a$ perturbations are independent, one can easily show that the last term in big bracket above vanishes. More specifically,
\ba
 \big \langle  \Delta \rho_v^{(\chi)} \Delta \rho_v^{(\phi) }  \big \rangle 
=  \langle  \Delta \rho_v^{(\chi)} \rangle  \langle  \Delta \rho_v^{(\phi)} \rangle
=0 \, . 
\ea
On the other hand, the fourth term in the big bracket is cubic in $\delta \phi$ perturbations. Since we assume that $\delta \phi$ perturbations are Gaussian, this contribution is suppressed compared to the first term. 

Now, noting that $\big \langle   (\Delta \rho_v)^2  \big\rangle $ for both $\phi$ and $\chi_a$ fields,   we obtain
\ba
\label{zeta-power}
\calP_\zeta\equiv 
\langle \zeta^2 \rangle &=& \big( \frac{H}{ \dot \rho^{\mathrm{cl}}_\phi}\big)^2
\Big(\big \langle   (\Delta \rho_\phi^{(1)})^2  \big \rangle + \delta \rho_\phi^2 +  \delta \rho_\chi^2 \Big) \nonumber\\
&=&
\calP_\zeta^{(0)} \Big(1+ \frac{\Delta \calP_\zeta}{\calP_\zeta^{(0)} }
\Big) \, .
\ea
Here $\calP_\zeta^{(0)}$ is the  the usual contribution in curvature perturbation power spectrum in the absence of zero point contributions, which is given by
\ba
\calP_\zeta^{(0)} = \big( \frac{H}{ \dot \rho^{\mathrm{cl}}_\phi}\big)^2
 \big \langle   (\Delta \rho_\phi^{(1)})^2  \big \rangle = \frac{H^2}{8 \pi^2 \epsilon_H M_P^2} \, .
\ea
Correspondingly, the fractional  correction  in power spectrum defined in Eq. (\ref{zeta-power}) is given by 
\ba 
\label{zeta-delta-power}
\frac{\Delta \calP_\zeta}{ \calP_\zeta^{(0)} } \equiv   \frac{\delta \rho_\phi^2 +  \delta \rho_\chi^2}{ \langle(\Delta \rho_\phi^{(1)})^2\rangle } \, .
\ea

From our previous discussions we have  $\delta \rho_\phi \propto m_\phi^4$
 while $\delta \rho_\chi \propto m_\chi^4$ with appropriate hierarchies among $\chi_a$ as discussed below Eq. (\ref{multiple-chi}). As we are interested in the spectroscopy of heavy spectator fields with $m_{\chi_a} > m_{\phi}$, we neglect the contribution of the inflaton zero point energy and,
\ba 
\label{zeta-delta-power2}
\frac{\Delta \calP_\zeta}{ \calP_\zeta^{(0)} } \simeq   
\frac{  \delta \rho_\chi^2}{ \langle(\Delta \rho_\phi^{(1)})^2\rangle } 
=  \frac{ \sum_a^{N-1}\delta \rho_{\chi_a}^2}{ \langle(\Delta \rho_\phi^{(1)})^2\rangle } \, .
\ea  
The implicit assumption here is that there are more than one heavy fields, i.e. 
$N\ge 3$. As discussed in previous section, if we have only one spectator field with $N=2$, then the normalization condition fixes $\langle \rho_\phi \rangle= -\langle \rho_\chi \rangle$ so $\delta \rho_\chi \sim m_\phi^4$ as 
seen in Eq. (\ref{light-cont}). Correspondingly, the fractional correction in power spectrum is not significant since $m_\phi \ll H$. Therefore, in order to have large correction in power spectrum, we assume that there are more than one heavy field in the spectrum with $N\ge3$.

Using the specific form of $\delta \rho_{\chi}^2$ from Eq. (\ref{delta-rho-tot})  
and the  formula for $\calP_\zeta^{(0)}$ from Eq. (\ref{zeta-delta-power}),  we obtain the following expression for the fractional correction in 
power spectrum, 
\ba
\label{frac}
\frac{\Delta \calP_\zeta}{ \calP_\zeta^{(0)} } \simeq  C_1
\calP_\zeta^{(0)}   \sum_{a, b} \Big[  \frac{m_a^4 m_b^4}{H^4 \sum_c m_c^4}   \ln \big(\frac{m_a^2}{m_b^2}\big) \Big]^2 \, ,
\ea
where $C_1$ is a numerical factor. It is understood that the sum over $a, b,c$ above is over the spectator fields $\chi_a$.

Note that we have calculated $\langle \zeta^2 \rangle$ in real space so there is no information of scale-dependence from Eq. (\ref{zeta-power}). To look for the  scale-dependence of the power spectrum, we should look at the scale-dependence of  various contributions of $\zeta$  in Eq. (\ref{zeta-flat2}) in Fourier space.  As shown in Appendix \ref{appendix},  the contributions of 
the vacuum zero point fluctuations  $\big \langle   (\Delta \rho_v^{(\phi)})^2  \big \rangle$ and  $ \big \langle   (\Delta \rho_v^{(\chi)})^2  \big\rangle$ have non-trivial scale-dependence in Fourier space, with $\Delta \calP_\zeta(k)$ typically scaling like $k^3$. This means that the correction in power spectrum associated to vacuum zero point energy mostly peaks on very small scales which are not observable in CMB observations. While these scales are observationally unaccessible, but their accumulative contribution in fractional power spectrum in real space is observable as given in  Eq. (\ref{frac}). More specifically, the amplitude of CMB perturbations is the total contribution  
$\calP_\zeta = \calP_\zeta^{(0)}+ \Delta \calP_\zeta$ which is fixed by COBE normalization to be $\calP_\zeta \simeq 2\times 10^{-9}$. Observationally, we can not separate $\calP_\zeta^{(0)}$ from 
$\Delta \calP_\zeta$ when looking at the amplitude of CMB perturbations. 
On the other hand, as we shall show in section \ref{loop}, the contribution from the vacuum zero pint fluctuations can be viewed as loop corrections in power spectrum. In order for the perturbative analysis to be under control, 
we  demand that the contribution from the vacuum zero point energy to be smaller than $\calP_\zeta^{(0)}$. This imposes the theoretical bound 
$\Delta \calP_\zeta < \calP_\zeta^{(0)} $. Using the specific form of $\Delta \calP_\zeta$ from Eq. (\ref{frac}) and neglecting the numerical prefactors of order unity,  this yields the following upper bound on the combination of the mass of the spectator fields,
\ba
\label{bound2}
{\cal M} \lesssim   {\calP_\zeta}^{\frac{-1}{8}} H \, ,
\ea
in which the mass scale ${\cal M}$ is defined via,
\ba
\label{M-mass}
{\cal M}^8 \equiv   \sum_{a, b} \Big(  \frac{m_a^4 m_b^4}{ \sum_c m_c^4}   \ln \big(\frac{m_a^2}{m_b^2}\big) \Big)^2 \, .
\ea

The  upper bound (\ref{bound2}) is indirect as it imposes a bound on the mass combination ${\cal M}$. As an example, consider the case of two heavy fields $N=3$ with $m_{\chi_1}\sim m_{\chi_2}\equiv m_\chi$. In this case, ${\cal M} \sim m_\chi $ and  the above bound is translated into 
\ba
m_\chi \lesssim {\calP_\zeta}^{\frac{-1}{8}} H \, .
\ea
Now consider the case of two heavy fields in which there is a large hierarchy amongst the masses of the heavy fields, say $m_{\chi_1}\ll  m_{\chi_2}$. 
In this case, ${\cal M} \sim m_{\chi_1} $ and 
the upper bound (\ref{bound2}) is translated into
\ba
m_{\chi_1} \lesssim {\calP_\zeta}^{\frac{-1}{8}} H \, ,
\ea
while the mass of the heaviest field $\chi_2$ is unconstrained. As we discussed in the previous section, this is a direct consequence of the renormalization condition. \\

\subsection{Bounds from Bispectrum}

To continue this line of investigation, now let us look at the non-Gaussianity induced by the fluctuations of the vacuum zero point energy. As we saw in 
Eq. (\ref{non-G}), the fluctuations of vacuum zero point energy associated to each field is highly non-Gaussian with $(\delta {\rho_v^{(i)}})^3\sim \langle 
\rho_v^{(i)} \rangle^3$. We expect this to induce large non-Gaussianity in curvature perturbations if the field is heavy.

The non-Gaussianity parameter $f_{NL}$ is roughly given by 
\cite{Chen:2010xka},
\ba
\label{fNL-eq}
f_{NL} \sim \frac{\langle \zeta^3 \rangle}{ \langle \zeta^2 \rangle^2 } \, .
\ea
The cosmological observations indicate that the primordial perturbations are nearly Gaussian with $|f_{NL}| \lesssim 1$ \cite{Planck:2019kim}.  On the other hand, the inflaton potential is nearly flat and its contribution in primordial non-Gaussianity is typically negligible \cite{Maldacena:2002vr}.  Therefore, any significant contribution in $f_{NL}$ comes from the fluctuations of
the vacuum zero point energy  of the heavy fields. Using Eq. (\ref{zeta-flat2}) for $\zeta$, the three-point function $\langle \zeta^3 \rangle$ is given by,
\ba
\label{zeta3}
\langle \zeta^3 \rangle = ( \frac{H}{ \dot \rho^{\mathrm{cl}}_\phi})^3
\big[  \langle   (\Delta \rho_\phi^{(1)})^3   \rangle  +  \langle   (\Delta \rho_v^{(\phi)})^3    +    (\Delta \rho_v^{(\chi)})^3 \rangle   + 3   \langle  \Delta \rho_\phi^{(1)} (\Delta \rho_v^{(\phi)})^2  \rangle + 3 \langle  (\Delta \rho_\phi^{(1)})^2 \Delta \rho_v^{(\phi)}   \rangle
 \big] 
\ea 
In obtaining  the above expression, we have used
$\langle  \Delta \rho_\phi^{(1)} \rangle= \langle  \Delta \rho_v^{(\phi)} \rangle
=  \langle  \Delta \rho_v^{(\chi)} \rangle=0$. 

The first term in Eq. (\ref{zeta3}) represents the non-Gaussianity associated to $\delta \phi$ perturbations. As we discussed before, this is very small in slow-roll limit \cite{Maldacena:2002vr} so we ignore its contribution in $f_{NL}$. The second and third terms  in Eq. (\ref{zeta3}) represent the skewness in vacuum zero point energy distribution as given in Eq. (\ref{non-G}). The fourth term containing $\langle  \Delta \rho_\phi^{(1)} (\Delta \rho_v^{(\phi)})^2  \rangle$ is odd in power of $\delta \phi$, at the order $\delta \phi^5$. Since $\delta \phi$ perturbations are Gaussian, the contribution from this term, like the first term,  is suppressed. Finally, the last term in Eq. (\ref{zeta3}) has fourth powers of $\delta \phi$ so it is not suppressed a priori. From the structure of this term, it will have the following form
\ba
\langle  (\Delta \rho_\phi^{(1)})^2 \Delta \rho_v^{(\phi)}   \rangle \sim 
\langle  (\Delta \rho_\phi^{(1)})^2 \rangle \langle \rho_v^{(\phi)} \rangle \, .
\ea
Now, we can compare the last term in Eq. (\ref{zeta3}) with the second term 
which is the contribution of the vacuum zero point fluctuations of inflaton, obtaining
\ba
\frac{\langle   (\Delta \rho_v^{(\phi)})^3\rangle}{\langle  (\Delta \rho_\phi^{(1)})^2 \Delta \rho_v^{(\phi)} \rangle} \sim 
\frac{\langle \rho_v^{(\phi)} \rangle^2}{\langle  (\Delta \rho_\phi^{(1)})^2 \rangle} \sim \calP_\zeta^{(0)} \big( \frac{m_\phi}{H}\big)^8 \, .
\ea
Since $m_\phi \ll H$,  we conclude that the contribution from the fluctuations of the vacuum zero point energy of inflaton  is much smaller than the last term in Eq. (\ref{zeta3}). Intuitively speaking, this is because the second term scales like $m_\phi^8$ while the last term scales like $m_\phi^4$. Since, $m_\phi \ll H$, we expect that the second term to be negligible compared to the last term.  

Now we calculate the contributions of the dominant terms, the third and the last terms of Eq. (\ref{zeta3}) in $f_{NL}$. Starting with the last term, and neglecting the numerical prefactors, its contribution in $f_{NL}$ is given by
\ba
f_{NL}|_{\mathrm{last \, term}}  \sim 
( \frac{H}{ \dot \rho^{\mathrm{cl}}_\phi})^3
 \langle  (\Delta \rho_\phi^{(1)})^2 \Delta \rho_v^{(\phi)}   \rangle 
 (\calP_\zeta^{(0)})^{-2} 
 \sim m_\phi^4  ( \calP_\zeta^{(0)})^{-2}\,    \calP_\zeta^{(0)}   ( \frac{H}{ \dot \rho^{\mathrm{cl}}_\phi}) \sim  \big(\frac{m_\phi}{H}\big)^4 \, .
\ea
As the inflaton field is light, we conclude that the above contribution in $f_{NL}$ is negligible. 
Therefore, the dominant contribution in $f_{NL}$ is entirely from the fluctuations of the vacuum zero point energy of the heavy spectator fields. 

Using our formula for skewness  Eq. (\ref{non-G}) and neglecting the contributions of the inflaton as discussed above, we obtain 
\ba
\label{fNL-bound}
f_{NL} &\simeq& ( \frac{H}{ \dot \rho^{\mathrm{cl}}_\phi})^3
 \langle  \big( \Delta \rho_v^{(\chi)} \big)^3  \rangle 
 (\calP_\zeta^{(0)})^{-2}    \sim  (\calP_\zeta^{(0)})^{-2}  \big( \frac{\calP_\zeta^{(0)}}{H^4}\big)^3 \sum_a \delta \rho_{\chi_a}^2 \nonumber\\
 &\simeq&  \calP_\zeta^{(0)} \frac{{\cal \tilde M}^{12}}{H^{12}}
\ea
in which the new mass scale $\cal \tilde M$ is defined via,
\ba
\label{tilde-M}
{\cal \tilde M}^{12} \equiv  \sum_{a, b} \Big(  \frac{m_a^4 m_b^4}{ \sum_c m_c^4}   \ln \big(\frac{m_i^2}{m_j^2}\big) \Big)^3 \, .
\ea
Note that the mass scales $\cal M$ and $ \cal \tilde M$ are similar but 
are not identical. 

In order to be consistent with cosmological observations with $|f_{NL}| \lesssim 1$, from Eq. (\ref{fNL-bound}) we obtain  the following upper bound on the combination of the masses of the heavy fields, 
\ba
\label{bound3}
{\cal \tilde M} \lesssim \calP_\zeta^{\frac{-1}{12}} H\, .
\ea
Numerically, this upper bound is similar to the upper bound 
(\ref{bound2}) obtained from the power spectrum. However, note that the upper bound (\ref{bound2}) is a theoretical requirement in order for the analysis to be under perturbative control while the bound (\ref{bound3}) is obtained from the observational constraint on non-Gaussianity. 

Again, the physical reason for this strong bound is that the inflaton perturbations are nearly Gaussian while the perturbations of the zero point energy are highly non-Gaussian. While the contributions of the zero point energy of the heavy fields  in the background expansion 
are negligible via our renormalization condition,  but their non-Gaussian properties are strong enough to affect the primordial bispectrum.  
Intuitively speaking, this situation is similar to the curvaton scenario. One can manage that the curvaton field to be subdominant in the background energy during inflation by a factor $R \ll1$. However, the perturbations become highly non-Gaussian with the amplitude $f_{NL} \sim 1/R$ \cite{Lyth:2001nq,  Sasaki:2006kq}. 

In conclusion, taking into account the uncertainties from the numerical prefactors, we conclude that the mass scales ${\cal \tilde M}, {\cal  M}$ associated to the fundamental  fields
can not be much heavier than  $H$, 
\ba
\label{bound-m}
{\cal \tilde M} , {\cal M} \lesssim H \, .
\ea
This is the main result of this work. 

This conclusion has important implications for physics beyond SM. For example,  from the upper bound $r < 10^{-2}$ on the amplitude of tensor to scalar spectra, 
we obtain the upper bound on the scale of inflation as $H\lesssim 10^{-5} M_P \sim 10^{13} \mathrm{GeV}$. Considering the numerical uncertainties of order unity in our analysis, this implies that the mass scales ${\cal \tilde M}, {\cal  M}$ in the  beyond SM sector should be lighter than $ 10^{14}\, \mathrm{GeV}$.  In general, in terms of the parameter $r$, we can express the upper bound (\ref{bound3}) as follows,
\ba
\label{bound-final}
{\cal \tilde M} \sim {\cal  M} \lesssim  \sqrt{r} \times 10^{14 }\mathrm{GeV} \, .
\ea
This implies that the mass scales ${\cal \tilde M}, {\cal  M}$ associated to the  fundamental fields are lighter than the GUT scale  by a factor $\sqrt{r}$. For example, if the scale of inflation happens to be very low, then these mass scales  are significantly below the GUT scale.  Eq. (\ref{bound-final}) sets upper bound on the combinations ${\cal \tilde M}$ and $ {\cal  M}$, but the direct bounds on the masses of the fundamental fields depend on the number of heavy fields and the relative hierarchy in the mass spectrum. For example, for the case of two heavy fields ($N=3$) with comparable masses of the order  
$m_\chi$, then ${\cal \tilde M}\sim {\cal  M} \sim  m_\chi$ and 
Eq. (\ref{bound-final}) imposes the bound $m_\chi < 10^{14}$ GeV.

While the upper bound  (\ref{bound-m}) is on the mass of the fundamental field, but it can be used to put bounds on the coupling of the heavy fields to the inflaton field as well. Suppose we have the interaction  ${\cal L}= \frac{1}{2}\sum_a g_a^2 \phi^2 \chi_a^2$. This will induce an effective mass $m_{\chi_a}$ for the field $\chi_a$ given by
\ba
m_{\chi_a}^2 = g_a^2 \phi^2 \, ,
\ea
in which $\phi$ is the classical value of the inflaton. Eq. (\ref{bound-m}) then can be used to impose upper bound on a combination of the couplings $g_a$. For example, if all couplings $g_a$ are the same order $g$, then from 
Eq. (\ref{bound-m}) we require that 
\ba
\label{g-bound}
g \lesssim \frac{H}{\phi}\, .
\ea
If the spectator fields $\chi_a$ is coupled to inflaton with couplings much stronger than the bound (\ref{g-bound}), then they induce a large mass for the spectator field, violating the bounds (\ref{bound-m}). As an example, suppose we have the large field model with $\phi \sim 10 M_P$ and 
$H \sim 10^{-5} M_P$.  
Then, the bound (\ref{g-bound}) requires $g\lesssim 10^{-6}$. \\


\subsection{Feynman Diagrams}
\label{loop}

While our analysis  were mostly based in real space, it is instructive to look at the corrections from vacuum zero point energy in Fourier space as well. To simplify the picture, we consider the case of a single heavy field $\chi$. 
Since the perturbations $ \Delta \rho_v^{(\chi)} $ is quadratic in $\chi^2$, 
the contribution of $ \Delta \rho_v^{(\chi)} $ in power spectrum of $\zeta_{\bf k}$ in Fourier space has the following form,
\ba
\label{power-k}
\langle \zeta_{\bf k_1}(\tau) \zeta_{\bf k_2}(\tau)\rangle \sim 
(2 \pi)^3 \delta^3({\bf k_1} + {\bf k_2}) 
( \frac{H}{ \dot \rho^{\mathrm{cl}}_\phi})^2
\int d^3 {\bf p} \,
\big|\chi_{\bf p}(\tau) \big|^2  \big|\chi_{{\bf k_1}- {\bf p} }(\tau) \big|^2  \, ,
\ea
where the symbol $\sim$ means we discard the numerical factors and other contributions such as $\dot \chi(\tau)^2$ and $(\nabla \chi)^2$.

As we demonstrate in Appendix \ref{appendix}, the mode functions are quite blue for massive fields.  The structure of the above convolution integral therefore suggests that, for a given mode $k$,   the contribution of the vacuum zero point energy  in $\calP_\zeta(k)$  comes from the UV modes in the integral over ${\bf p}$.  Here $\tau$ is any representative time when the mode of interest $k$ has left the horizon and the leading contribution in $\zeta$, i.e. the first term in Eq. (\ref{zeta-flat}), freezes. This can be a few e-folds after the time of horizon crossing for the mode $k$
or simply the time of end of inflation. As the variance $\delta \rho_v^2$ and skewness $\delta \rho_v^3$ are constant (independent of time), the above correlation can be calculated at any time as long as $\zeta$ freezes.

Similarly, the contribution of $ \Delta \rho_v^{(\chi)} $ in bispectrum has the following form
\ba
\label{bispectrum-k}
\langle \zeta_{\bf k_1} \zeta_{\bf k_2} \zeta_{\bf k_3}
\rangle \sim 
(2 \pi)^3 \delta^3({\bf k_1} + {\bf k_2} + {\bf k_3}) 
( \frac{H}{ \dot \rho^{\mathrm{cl}}_\phi})^3
\int d^3 {\bf p} 
\big|\chi_{\bf p}(\tau) \big|^2  \big|\chi_{{\bf k_2}+ {\bf p} }(\tau) \big|^2  \big|\chi_{{\bf k_1}- {\bf p} }(\tau) \big|^2  .
\ea


\begin{figure}[t]
\vspace{-0.5cm}
\begin{center}
	\includegraphics[scale=0.7]{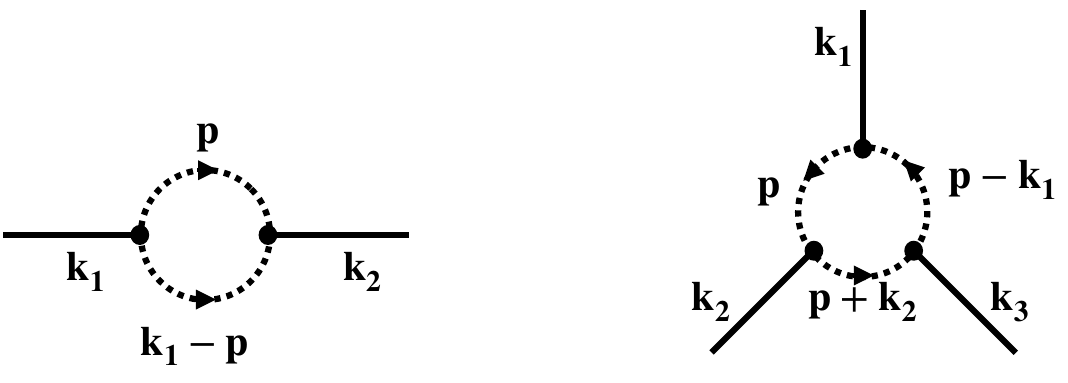}
	\end{center}
\caption{ The Feynman diagrams for the contributions of the vacuum zero point fluctuations of the heavy fields 
in power spectrum (left) and bispectrum (right). The external solid  lines represent  $\zeta_{\bf k_i}$ while the dotted curves represent the propagator corresponding to $|\chi_{\bf p}|^2, |\chi_{ {\bf k_1} -{\bf p}}|^2$ etc. Each vertex has the  amplitude  $( \frac{H}{ \dot \rho^{\mathrm{cl}}_\phi})$. 
See also \cite{Byrnes:2007tm} for similar Feynman diagrams.  
\label{Feynman}
}
\end{figure}

It would be instructive to look at the above results in terms of Feynman diagrams. In Figure \ref{Feynman} we have presented the Feynman diagrams for the contributions of the fluctuations of the vacuum zero point energy from the heavy field in power spectrum and bispectrum. The structure of the integrals in Eqs. (\ref{power-k}) and (\ref{bispectrum-k}) indicates that these contributions are in the form of one-loop corrections. The small scale modes that are running inside the loops yield the dominant contributions in the integrals in  Eqs. (\ref{power-k}) and (\ref{bispectrum-k}). As the correction in power spectrum $\Delta \calP_\zeta(k)$ is blue, the long CMB scale modes are unaffected from the loop corrections. Instead, the corrections in power spectrum is significant on small scales. In this view, the effects of one-loop corrections here are different than  the one-loop corrections in \cite{Kristiano:2022maq, Kristiano:2024ngc, Riotto:2023hoz, Choudhury:2023vuj, Choudhury:2023jlt, Firouzjahi:2023ahg, Firouzjahi:2023aum} where it is shown  that short  modes which experience  an intermediate phase of ultra slow-roll  inflation can affect the long CMB scale mode. More specifically, $\Delta \calP_\zeta(k)$ from the loop corrections in the latter setup is scale-invariant so  the long CMB scale modes and the  short modes are affected similarly.  However, in our case $\Delta \calP_\zeta(k)$ has a strong blue scale-dependence so the long modes are protected from large loop corrections. 

Before closing this section we comment that the roles of the heavy spectator fields were investigated  in \cite{Chen:2012ye}.\footnote{
We thank Xingang Chen for bringing \cite{Chen:2012ye} to our attention 
while our work was in its final stage.} In that work the authors used perturbative in-in formalism to calculate the corrections in power spectrum from quartic interactions of the type $m^2 \zeta^2 \chi^2$. To regularize the UV divergent integrals, they imposed a cut-off $\Lambda$ by hand obtaining a correction of the form $\Delta \calP_\zeta/\calP_\zeta \sim \calP_\zeta^{(0)} (\Lambda/H)^4$. Comparing their result with  our result Eq. (\ref{frac}), there are two important differences. First, we do not have the cutoff $\Lambda$ as we perform the regularization automatically via DR scheme. In a sense, their $\Lambda$ will be replaced by the mass of the field $m$. Second, their fractional correction in power spectrum scales like $\Lambda^4/H^4$ while ours scales like $m^8/H^8$. The reason is that they used the quartic Hamiltonian of the type $m^2 \zeta^2 \chi^2$. To obtain our scaling  $m^8/H^8$, one should start with a cubic Hamiltonian of the form $\zeta \chi^2$ in the analysis of \cite{Chen:2012ye} which yields to a Feynman diagram similar to the left panel of Fig. \ref{Feynman} with a nested in-in integral. Since our result for $\Delta \calP_\zeta$ is expressed in term of $m^8$ we are able to put an upper bound on the mass of field while in \cite{Chen:2012ye} the bound will be imposed on $\Lambda$ which was interpreted as the scale of the UV completed theory. \\

\section{Summary and Discussions}
\label{summary}

In this work we have studied the implications from the fluctuations of the vacuum zero point energy associated to fundamental fields during inflation. 
At the background level, the vacuum zero point energy associated to a field with mass $m$ contributes to the cosmological constant of the order $m^4$.
This is the source of the infamous cosmological constant problem. There is no compelling dynamical mechanism to tune the contributions of the quantum fields in cosmological constant to be consistent with the magnitude of dark energy as observed in  cosmological observations. One may simply set the cosmological constant induced by quantum fields to be zero (or very nearly zero) at the background level. However, the crucial observation is  that the perturbations in the distribution of the vacuum zero point energy scales like the background vacuum energy, i.e. $\delta \rho_v \sim m^4$. While one may 
absorb  the background vacuum zero point energy by some mechanism, however the perturbations in distribution of vacuum energy are always present. This shows  another face of the cosmological constant problem, now at the level of perturbations. We comment that the inhomogeneities in the distribution of the vacuum zero point energy  
were also advocated by Unruh and his collaborators in  \cite{Wang:2017oiy, Cree:2018mcx, Wang:2019mbh, Wang:2019mee, Wang:2023tzm}. 

A key step in dealing with the effects of vacuum zero point energy is the condition of renormalization to get rid of the mass parameter $\mu$. Our renormalization condition, motivated by cosmological observation, is that the total vacuum zero point energy induced by all fundamental fields to be zero,
$\langle \rho_v\rangle_{\mathrm{tot}}=0$. This in turn fixes $\mu$ in terms of the masses involved in the theory such that  $\langle \rho_i\rangle$ associated to each field is fixed by Eq. (\ref{rho-i}).

The  fluctuations in vacuum zero point energy contribute to primordial curvature perturbations.  We have shown that in order to keep the primordial perturbations to remain nearly  Gaussian, the mass scale ${\cal \tilde M}$ associated to the  fundamental fields can not be significantly heavier than $H$. This is a  strong conclusion. A similar upper bound on the mass scale 
${\cal  M}$ is obtained from the theoretical requirement that the loop corrections in power spectrum to be under perturbative control.   In terms of the parameter $r$, our bound is translated into ${\cal \tilde M}, {\cal M} \lesssim \sqrt{r} \times 10^{14}  \mathrm{GeV}$. This conclusion has important implications for physics beyond SM. For example,  consider a setup of two heavy spectator fields of comparable masses. Then from the upper bound $r < 10^{-2}$, and considering numerical uncertainties of order unity in our analysis,  
we conclude that  the fields should be lighter than $10^{14}\mathrm{GeV}$. This is just below  the GUT scale. 

While we presented the  analysis for spectator scalar fields, but the result can be extended to other fields with different spins as well. For example, as shown in \cite{Firouzjahi:2022xxb}, the fluctuations in vacuum zero point energy of the fermionic fields also satisfy the relation $\delta \rho_v \sim \langle \rho_v \rangle \sim m^4$. Since this relation was the key ingredient in the derivation of our upper bound on the mass, we conclude that our upper bound applies to fermionic fields as well. This conclusion  applies for massive gauge bosons with spin one as well. 

In terms of the Feynman diagrams, the corrections from the vacuum zero point fluctuations can be interpreted as one-loop corrections in power spectrum and bispectrum. Since the spectrum of $ \Delta \rho_v^{(\chi)} $ 
is highly blue,  the leading contributions from these loop corrections come from small scale modes which run inside the loop. Since the correction in power spectrum $\Delta \calP_\zeta(k)$ is blue-tilted, the loop corrections affect the short scales while the long modes, such as the CMB scale modes, are largely unaffected by these quantum loop corrections.

\vspace{0.7cm}

{\bf Acknowledgments:}  We thank Yashar Akrami, Robert Brandenberger, 
Xingang Chen, Mohammad Hossein Namjoo, Misao Sasaki and Haidar Sheikhahmadi  for useful discussions and correspondences.  We are grateful to the anonymous referee for the insightful comments which have improved our 
studies.  This work is supported by the INSF  of Iran under the grant  number 4022911.

\appendix
\section{Scale-dependence of $\Delta \rho_v^{(\phi)} $ and $ \Delta \rho_v^{(\chi)}$ in $\zeta$}
\label{appendix}

In this appendix we investigate the scale-dependence of the contributions 
from the perturbations in vacuum zero point fluctuations $\Delta \rho_v^{(\phi)} $ and $ \Delta \rho_v^{(\chi)}$ in $\zeta$.
Our goal is to show that since these contributions are quadratic in field perturbations, respectively $\delta \phi^2$ and $\delta \chi^2$, then their contributions are highly scale-dependent in Fourier space.

The curvature perturbation on the surface of constant energy density is given 
by 
\ba
\label{zeta-flat3}
\zeta=  \frac{H}{ \dot \rho^{\mathrm{cl}}_\phi} 
\big( \Delta \rho_\phi^{(1)} + \Delta \rho_v^{(\phi)} + 
\Delta \rho_v^{(\chi)}\big) \, .
\ea
The first term above is the usual scale-invariant term. To see this, let us look 
at the mode functions, 
\begin{equation}
\label{chi-k2}
\delta \phi_k(\tau)\, ,\chi _k(\tau)  = {( - H\tau )^{\frac{{D - 1}}{2}}}{\left( {\frac{\pi }{{4H}}} \right)^{\frac{1}{2}}} {{e^{\frac{i \pi}{2}  (\nu + \frac{1}{2})} }}
H_\nu ^{(1)}( - k\tau ){\mkern 1mu}\,,
\end{equation}
where
\begin{equation}
\label{nu-002}
{{
\nu  \equiv  \frac{1}{2}{\mkern 1mu} \sqrt {9- 4 \beta^2}\, , \quad \quad
\beta \equiv \frac{m}{H} }}\, .
\end{equation}
For inflaton field, $\beta \ll 1$ so $\nu \simeq \frac{3}{2}$. However, for the spectator field $\chi$, if $\beta \sim1$ then $\nu$ is far from the critical 
value $\frac{3}{2}$ while for larger values of $\beta$ it can even be a complex number. 

Now let us look at the superhorizon limit where $k \tau \rightarrow 0$. In a sense, we calculate the power spectrum at the end of inflation $\tau \rightarrow 0$ so all modes of interests are superhorizon. Using the small argument limit of the Hankel function, we have (assuming $\nu$ is real)
\ba
H_\nu ^{(1)}( - k\tau ) \simeq -\frac{i}{\pi} \Gamma(\nu) \big(  \frac{-k \tau}{2} \big)^{-\nu} \,.
\ea
We see that the mode function scales like $(- k \tau)^{-\nu}$ on superhorizon scales. As the dimensionless power spectrum $\calP_\zeta$ is defined via
\ba
\label{calP}
\calP_\zeta \equiv \frac{k^3}{2 \pi^2} |\zeta_k|^2 \, ,
\ea
we conclude that the first term in Eq. (\ref{zeta-flat3}) which is linear in 
$\delta \phi_k$ scales like $k^3 k^{-2\nu} = k^{3- 2 \nu}$. Since for inflaton 
$\nu \simeq \frac{3}{2}$, the power spectrum is nearly scale-invariant. The deviation in scale-invariance is determined by the slow-roll corrections.

Now we investigate the scale-dependence of the remaining two terms in Eq. (\ref{zeta-flat3}). As both of them have similar forms, we consider  the third term induced from the spectator field. Since $\Delta \rho_v^{(\chi)}  \sim  \chi^2$, its contribution in power spectrum of $\zeta_\bfk$ in Fourier space has the following form, 
\ba
\label{power-k-b}
\langle \zeta_{\bf k_1}(\tau) \zeta_{\bf k_2}(\tau)\rangle \sim 
(2 \pi)^3 \delta^3({\bf k_1} + {\bf k_2}) 
( \frac{H}{ \dot \rho^{\mathrm{cl}}_\phi})^2
\int d^3 {\bf p} \,
\big|\chi_{\bf p}(\tau) \big|^2  \big|\chi_{{\bf k_1}- {\bf p} }(\tau) \big|^2  \, ,
\ea
where, as mentioned in the main text,  the symbol $\sim$ means we discard the numerical factors and other contributions such as $\dot \chi(\tau)^2$ and $(\nabla \chi)^2$.  As the integral in Eq. (\ref{power-k-b}) is UV divergent, we expect the dominant contribution to come from the modes deep inside the horizon, i.e. from modes which experience  the flat Minkowski background
with $p \rightarrow \infty$.  In this limit $\chi_p \sim p^{-1/2}$  so 
\ba
\label{super-int}
\langle \zeta_{\bf k_1}(\tau) \zeta_{\bf k_2}(\tau)\rangle  \sim \int d^3 {\bf p} \frac{1}{p^2} \, .
\ea
As expected, the above integral is UV divergent which is the hallmark of the vacuum zero point energy and its fluctuations.  After regularizing this divergence (as we did via DR in  section (\ref{vac-reg})), we conclude that $\langle \zeta_{\bf k_1} \zeta_{\bf k_2}\rangle$ is nearly independent of $k$. Constructing 
$\calP_\zeta\sim k^3 \langle \zeta_{\bf k_1} \zeta_{\bf k_2}\rangle$ we conclude that $\calP_\zeta$ is blue scaling like $k^3$.


\bibliography{references2}{}

\providecommand{\href}[2]{#2}\begingroup\raggedright\begin{thebibliography}{10}

\bibitem{Casimir:1948dh}
H.~B.~G. Casimir\emph{}; {\emph{Indag. Math.} {\bfseries 10} (1948) 261--263}.

\bibitem{Lamoreaux:1996wh}
S.~K. Lamoreaux\emph{};
  \href{https://doi.org/10.1103/PhysRevLett.78.5}{\emph{Phys. Rev. Lett.}
  {\bfseries 78} (1997) 5--8}.

\bibitem{Bordag:2001qi}
M.~Bordag, U.~Mohideen and V.~M. Mostepanenko\emph{};
  \href{https://doi.org/10.1016/S0370-1573(01)00015-1}{\emph{Phys. Rept.}
  {\bfseries 353} (2001) 1--205},
  [\href{https://arxiv.org/abs/quant-ph/0106045}{{\ttfamily
  quant-ph/0106045}}].

\bibitem{Planck:2018vyg}
{\scshape Planck} collaboration, N.~Aghanim et~al.\emph{};
  \href{https://doi.org/10.1051/0004-6361/201833910}{\emph{Astron. Astrophys.}
  {\bfseries 641} (2020) A6},
  [\href{https://arxiv.org/abs/1807.06209}{{\ttfamily 1807.06209}}].

\bibitem{SupernovaCosmologyProject:1998vns}
{\scshape Supernova Cosmology Project} collaboration, S.~Perlmutter
  et~al.\emph{}; \href{https://doi.org/10.1086/307221}{\emph{Astrophys. J.}
  {\bfseries 517} (1999) 565--586},
  [\href{https://arxiv.org/abs/astro-ph/9812133}{{\ttfamily
  astro-ph/9812133}}].

\bibitem{SupernovaSearchTeam:1998fmf}
{\scshape Supernova Search Team} collaboration, A.~G. Riess et~al.\emph{};
  \href{https://doi.org/10.1086/300499}{\emph{Astron. J.} {\bfseries 116}
  (1998) 1009--1038}, [\href{https://arxiv.org/abs/astro-ph/9805201}{{\ttfamily
  astro-ph/9805201}}].

\bibitem{Weinberg:1988cp}
S.~Weinberg\emph{}; \href{https://doi.org/10.1103/RevModPhys.61.1}{\emph{Rev.
  Mod. Phys.} {\bfseries 61} (1989) 1--23}.

\bibitem{Sahni:1999gb}
V.~Sahni and A.~A. Starobinsky\emph{};
  \href{https://doi.org/10.1142/S0218271800000542}{\emph{Int. J. Mod. Phys. D}
  {\bfseries 9} (2000) 373--444},
  [\href{https://arxiv.org/abs/astro-ph/9904398}{{\ttfamily
  astro-ph/9904398}}].

\bibitem{Peebles:2002gy}
P.~J.~E. Peebles and B.~Ratra\emph{};
  \href{https://doi.org/10.1103/RevModPhys.75.559}{\emph{Rev. Mod. Phys.}
  {\bfseries 75} (2003) 559--606},
  [\href{https://arxiv.org/abs/astro-ph/0207347}{{\ttfamily
  astro-ph/0207347}}].

\bibitem{Copeland:2006wr}
E.~J. Copeland, M.~Sami and S.~Tsujikawa\emph{};
  \href{https://doi.org/10.1142/S021827180600942X}{\emph{Int. J. Mod. Phys. D}
  {\bfseries 15} (2006) 1753--1936},
  [\href{https://arxiv.org/abs/hep-th/0603057}{{\ttfamily hep-th/0603057}}].

\bibitem{Christensen:1976vb}
S.~M. Christensen\emph{};
  \href{https://doi.org/10.1103/PhysRevD.14.2490}{\emph{Phys. Rev. D}
  {\bfseries 14} (1976) 2490--2501}.

\bibitem{Christensen:1977jc}
S.~M. Christensen and S.~A. Fulling\emph{};
  \href{https://doi.org/10.1103/PhysRevD.15.2088}{\emph{Phys. Rev. D}
  {\bfseries 15} (1977) 2088--2104}.

\bibitem{Davies:1977ze}
P.~C.~W. Davies, S.~A. Fulling, S.~M. Christensen and T.~S. Bunch\emph{};
  \href{https://doi.org/10.1016/0003-4916(77)90167-1}{\emph{Annals Phys.}
  {\bfseries 109} (1977) 108--142}.

\bibitem{Anderson:1990jh}
P.~R. Anderson\emph{};
  \href{https://doi.org/10.1103/PhysRevD.41.1152}{\emph{Phys. Rev. D}
  {\bfseries 41} (1990) 1152--1162}.

\bibitem{tHooft:1972tcz}
G.~'t~Hooft and M.~J.~G. Veltman\emph{};
  \href{https://doi.org/10.1016/0550-3213(72)90279-9}{\emph{Nucl. Phys. B}
  {\bfseries 44} (1972) 189--213}.

\bibitem{tHooft:1974toh}
G.~'t~Hooft and M.~J.~G. Veltman\emph{}; {\emph{Ann. Inst. H. Poincare Phys.
  Theor. A} {\bfseries 20} (1974) 69--94}.

\bibitem{Bollini:1972ui}
C.~G. Bollini and J.~J. Giambiagi\emph{};
  \href{https://doi.org/10.1007/BF02895558}{\emph{Nuovo Cim. B} {\bfseries 12}
  (1972) 20--26}.

\bibitem{Deser:1974cz}
S.~Deser and P.~van Nieuwenhuizen\emph{};
  \href{https://doi.org/10.1103/PhysRevD.10.401}{\emph{Phys. Rev. D} {\bfseries
  10} (1974) 401}.

\bibitem{Dowker:1975tf}
J.~S. Dowker and R.~Critchley\emph{};
  \href{https://doi.org/10.1103/PhysRevD.13.3224}{\emph{Phys. Rev. D}
  {\bfseries 13} (1976) 3224}.

\bibitem{Barvinsky:1985an}
A.~O. Barvinsky and G.~A. Vilkovisky\emph{};
  \href{https://doi.org/10.1016/0370-1573(85)90148-6}{\emph{Phys. Rept.}
  {\bfseries 119} (1985) 1--74}.

\bibitem{Onemli:2002hr}
V.~K. Onemli and R.~P. Woodard\emph{};
  \href{https://doi.org/10.1088/0264-9381/19/17/311}{\emph{Class. Quant. Grav.}
  {\bfseries 19} (2002) 4607},
  [\href{https://arxiv.org/abs/gr-qc/0204065}{{\ttfamily gr-qc/0204065}}].

\bibitem{Brunier:2004sb}
T.~Brunier, V.~K. Onemli and R.~P. Woodard\emph{};
  \href{https://doi.org/10.1088/0264-9381/22/1/005}{\emph{Class. Quant. Grav.}
  {\bfseries 22} (2005) 59--84},
  [\href{https://arxiv.org/abs/gr-qc/0408080}{{\ttfamily gr-qc/0408080}}].

\bibitem{Miao:2005am}
S.-P. Miao and R.~P. Woodard\emph{};
  \href{https://doi.org/10.1088/0264-9381/23/5/016}{\emph{Class. Quant. Grav.}
  {\bfseries 23} (2006) 1721--1762},
  [\href{https://arxiv.org/abs/gr-qc/0511140}{{\ttfamily gr-qc/0511140}}].

\bibitem{Prokopec:2008gw}
T.~Prokopec, N.~C. Tsamis and R.~P. Woodard\emph{};
  \href{https://doi.org/10.1103/PhysRevD.78.043523}{\emph{Phys. Rev. D}
  {\bfseries 78} (2008) 043523},
  [\href{https://arxiv.org/abs/0802.3673}{{\ttfamily 0802.3673}}].

\bibitem{Miao:2010vs}
S.~P. Miao, N.~C. Tsamis and R.~P. Woodard\emph{};
  \href{https://doi.org/10.1063/1.3448926}{\emph{J. Math. Phys.} {\bfseries 51}
  (2010) 072503}, [\href{https://arxiv.org/abs/1002.4037}{{\ttfamily
  1002.4037}}].

\bibitem{Glavan:2021adm}
D.~Glavan, S.~P. Miao, T.~Prokopec and R.~P. Woodard\emph{};
  \href{https://doi.org/10.1007/JHEP03(2022)088}{\emph{JHEP} {\bfseries 03}
  (2022) 088}, [\href{https://arxiv.org/abs/2112.00959}{{\ttfamily
  2112.00959}}].

\bibitem{Glavan:2020gal}
D.~Glavan, S.~P. Miao, T.~Prokopec and R.~P. Woodard\emph{};
  \href{https://doi.org/10.1103/PhysRevD.101.106016}{\emph{Phys. Rev. D}
  {\bfseries 101} (2020) 106016},
  [\href{https://arxiv.org/abs/2003.02549}{{\ttfamily 2003.02549}}].

\bibitem{Martin:2012bt}
J.~Martin\emph{};
  \href{https://doi.org/10.1016/j.crhy.2012.04.008}{\emph{Comptes Rendus
  Physique} {\bfseries 13} (2012) 566--665},
  [\href{https://arxiv.org/abs/1205.3365}{{\ttfamily 1205.3365}}].

\bibitem{Akhmedov:2002ts}
E.~K. Akhmedov\emph{};  \href{https://arxiv.org/abs/hep-th/0204048}{{\ttfamily
  hep-th/0204048}}.

\bibitem{Ossola:2003ku}
G.~Ossola and A.~Sirlin\emph{};
  \href{https://doi.org/10.1140/epjc/s2003-01337-7}{\emph{Eur. Phys. J. C}
  {\bfseries 31} (2003) 165--175},
  [\href{https://arxiv.org/abs/hep-ph/0305050}{{\ttfamily hep-ph/0305050}}].

\bibitem{Hawking:1975vcx}
S.~W. Hawking\emph{}; \href{https://doi.org/10.1007/BF02345020}{\emph{Commun.
  Math. Phys.} {\bfseries 43} (1975) 199--220}.

\bibitem{Unruh:1976db}
W.~G. Unruh\emph{}; \href{https://doi.org/10.1103/PhysRevD.14.870}{\emph{Phys.
  Rev. D} {\bfseries 14} (1976) 870}.

\bibitem{Unruh:1983ms}
W.~G. Unruh and R.~M. Wald\emph{};
  \href{https://doi.org/10.1103/PhysRevD.29.1047}{\emph{Phys. Rev. D}
  {\bfseries 29} (1984) 1047--1056}.

\bibitem{Jacobson:2003vx}
T.~Jacobson\emph{};  in \emph{{School on Quantum Gravity}}, pp.~39--89, 8,
  2003, \href{https://arxiv.org/abs/gr-qc/0308048}{{\ttfamily gr-qc/0308048}},
  \href{https://doi.org/10.1007/0-387-24992-3_2}{DOI}.

\bibitem{Firouzjahi:2022rtn}
H.~Firouzjahi and A.~Talebian\emph{};
  \href{https://doi.org/10.1103/PhysRevD.106.123505}{\emph{Phys. Rev. D}
  {\bfseries 106} (2022) 123505},
  [\href{https://arxiv.org/abs/2210.15186}{{\ttfamily 2210.15186}}].

\bibitem{Onemli:2004mb}
V.~K. Onemli and R.~P. Woodard\emph{};
  \href{https://doi.org/10.1103/PhysRevD.70.107301}{\emph{Phys. Rev. D}
  {\bfseries 70} (2004) 107301},
  [\href{https://arxiv.org/abs/gr-qc/0406098}{{\ttfamily gr-qc/0406098}}].

\bibitem{Janssen:2008px}
T.~M. Janssen, S.~P. Miao, T.~Prokopec and R.~P. Woodard\emph{};
  \href{https://doi.org/10.1088/0264-9381/25/24/245013}{\emph{Class. Quant.
  Grav.} {\bfseries 25} (2008) 245013},
  [\href{https://arxiv.org/abs/0808.2449}{{\ttfamily 0808.2449}}].

\bibitem{Firouzjahi:2022xxb}
H.~Firouzjahi\emph{};
  \href{https://doi.org/10.1103/PhysRevD.106.083510}{\emph{Phys. Rev. D}
  {\bfseries 106} (2022) 083510},
  [\href{https://arxiv.org/abs/2201.02016}{{\ttfamily 2201.02016}}].

\bibitem{Firouzjahi:2023wbe}
H.~Firouzjahi and H.~Sheikhahmadi\emph{};
  \href{https://doi.org/10.1103/PhysRevD.108.065002}{\emph{Phys. Rev. D}
  {\bfseries 108} (2023) 065002},
  [\href{https://arxiv.org/abs/2307.00977}{{\ttfamily 2307.00977}}].

\bibitem{Firouzjahi:2022vij}
H.~Firouzjahi\emph{};
  \href{https://doi.org/10.1103/PhysRevD.106.045015}{\emph{Phys. Rev. D}
  {\bfseries 106} (2022) 045015},
  [\href{https://arxiv.org/abs/2205.06561}{{\ttfamily 2205.06561}}].

\bibitem{Arkani-Hamed:2015bza}
N.~Arkani-Hamed and J.~Maldacena\emph{};
  \href{https://arxiv.org/abs/1503.08043}{{\ttfamily 1503.08043}}.

\bibitem{Meerburg:2016zdz}
P.~D. Meerburg, M.~M\"unchmeyer, J.~B. Mu\~noz and X.~Chen\emph{};
  \href{https://doi.org/10.1088/1475-7516/2017/03/050}{\emph{JCAP} {\bfseries
  03} (2017) 050}, [\href{https://arxiv.org/abs/1610.06559}{{\ttfamily
  1610.06559}}].

\bibitem{Chen:2016uwp}
X.~Chen, Y.~Wang and Z.-Z. Xianyu\emph{};
  \href{https://doi.org/10.1103/PhysRevLett.118.261302}{\emph{Phys. Rev. Lett.}
  {\bfseries 118} (2017) 261302},
  [\href{https://arxiv.org/abs/1610.06597}{{\ttfamily 1610.06597}}].

\bibitem{Wang:2019gbi}
L.-T. Wang and Z.-Z. Xianyu\emph{};
  \href{https://doi.org/10.1007/JHEP02(2020)044}{\emph{JHEP} {\bfseries 02}
  (2020) 044}, [\href{https://arxiv.org/abs/1910.12876}{{\ttfamily
  1910.12876}}].

\bibitem{Chen:2009zp}
X.~Chen and Y.~Wang\emph{};
  \href{https://doi.org/10.1088/1475-7516/2010/04/027}{\emph{JCAP} {\bfseries
  04} (2010) 027}, [\href{https://arxiv.org/abs/0911.3380}{{\ttfamily
  0911.3380}}].

\bibitem{Noumi:2012vr}
T.~Noumi, M.~Yamaguchi and D.~Yokoyama\emph{};
  \href{https://doi.org/10.1007/JHEP06(2013)051}{\emph{JHEP} {\bfseries 06}
  (2013) 051}, [\href{https://arxiv.org/abs/1211.1624}{{\ttfamily 1211.1624}}].

\bibitem{Emami:2013lma}
R.~Emami\emph{};
  \href{https://doi.org/10.1088/1475-7516/2014/04/031}{\emph{JCAP} {\bfseries
  04} (2014) 031}, [\href{https://arxiv.org/abs/1311.0184}{{\ttfamily
  1311.0184}}].

\bibitem{Wang:2017oiy}
Q.~Wang, Z.~Zhu and W.~G. Unruh\emph{};
  \href{https://doi.org/10.1103/PhysRevD.95.103504}{\emph{Phys. Rev. D}
  {\bfseries 95} (2017) 103504},
  [\href{https://arxiv.org/abs/1703.00543}{{\ttfamily 1703.00543}}].

\bibitem{Cree:2018mcx}
S.~S. Cree, T.~M. Davis, T.~C. Ralph, Q.~Wang, Z.~Zhu and W.~G. Unruh\emph{};
  \href{https://doi.org/10.1103/PhysRevD.98.063506}{\emph{Phys. Rev. D}
  {\bfseries 98} (2018) 063506},
  [\href{https://arxiv.org/abs/1805.12293}{{\ttfamily 1805.12293}}].

\bibitem{Wang:2019mbh}
Q.~Wang and W.~G. Unruh\emph{};
  \href{https://doi.org/10.1103/PhysRevD.102.023537}{\emph{Phys. Rev. D}
  {\bfseries 102} (2020) 023537},
  [\href{https://arxiv.org/abs/1904.08599}{{\ttfamily 1904.08599}}].

\bibitem{Wang:2019mee}
Q.~Wang\emph{};
  \href{https://doi.org/10.1103/PhysRevLett.125.051301}{\emph{Phys. Rev. Lett.}
  {\bfseries 125} (2020) 051301},
  [\href{https://arxiv.org/abs/1904.09566}{{\ttfamily 1904.09566}}].

\bibitem{Wang:2023tzm}
Q.~Wang\emph{};  \href{https://arxiv.org/abs/2312.06692}{{\ttfamily
  2312.06692}}.

\bibitem{Lyth:2004gb}
D.~H. Lyth, K.~A. Malik and M.~Sasaki\emph{};
  \href{https://doi.org/10.1088/1475-7516/2005/05/004}{\emph{JCAP} {\bfseries
  05} (2005) 004}, [\href{https://arxiv.org/abs/astro-ph/0411220}{{\ttfamily
  astro-ph/0411220}}].

\bibitem{Bassett:2005xm}
B.~A. Bassett, S.~Tsujikawa and D.~Wands\emph{};
  \href{https://doi.org/10.1103/RevModPhys.78.537}{\emph{Rev. Mod. Phys.}
  {\bfseries 78} (2006) 537--589},
  [\href{https://arxiv.org/abs/astro-ph/0507632}{{\ttfamily
  astro-ph/0507632}}].

\bibitem{Planck:2018jri}
{\scshape Planck} collaboration, Y.~Akrami et~al.\emph{};
  \href{https://doi.org/10.1051/0004-6361/201833887}{\emph{Astron. Astrophys.}
  {\bfseries 641} (2020) A10},
  [\href{https://arxiv.org/abs/1807.06211}{{\ttfamily 1807.06211}}].

\bibitem{Chen:2010xka}
X.~Chen\emph{}; \href{https://doi.org/10.1155/2010/638979}{\emph{Adv. Astron.}
  {\bfseries 2010} (2010) 638979},
  [\href{https://arxiv.org/abs/1002.1416}{{\ttfamily 1002.1416}}].

\bibitem{Planck:2019kim}
{\scshape Planck} collaboration, Y.~Akrami et~al.\emph{};
  \href{https://doi.org/10.1051/0004-6361/201935891}{\emph{Astron. Astrophys.}
  {\bfseries 641} (2020) A9},
  [\href{https://arxiv.org/abs/1905.05697}{{\ttfamily 1905.05697}}].

\bibitem{Maldacena:2002vr}
J.~M. Maldacena\emph{};
  \href{https://doi.org/10.1088/1126-6708/2003/05/013}{\emph{JHEP} {\bfseries
  05} (2003) 013}, [\href{https://arxiv.org/abs/astro-ph/0210603}{{\ttfamily
  astro-ph/0210603}}].

\bibitem{Lyth:2001nq}
D.~H. Lyth and D.~Wands\emph{};
  \href{https://doi.org/10.1016/S0370-2693(01)01366-1}{\emph{Phys. Lett. B}
  {\bfseries 524} (2002) 5--14},
  [\href{https://arxiv.org/abs/hep-ph/0110002}{{\ttfamily hep-ph/0110002}}].

\bibitem{Sasaki:2006kq}
M.~Sasaki, J.~Valiviita and D.~Wands\emph{};
  \href{https://doi.org/10.1103/PhysRevD.74.103003}{\emph{Phys. Rev. D}
  {\bfseries 74} (2006) 103003},
  [\href{https://arxiv.org/abs/astro-ph/0607627}{{\ttfamily
  astro-ph/0607627}}].

\bibitem{Byrnes:2007tm}
C.~T. Byrnes, K.~Koyama, M.~Sasaki and D.~Wands\emph{};
  \href{https://doi.org/10.1088/1475-7516/2007/11/027}{\emph{JCAP} {\bfseries
  11} (2007) 027}, [\href{https://arxiv.org/abs/0705.4096}{{\ttfamily
  0705.4096}}].

\bibitem{Kristiano:2022maq}
J.~Kristiano and J.~Yokoyama\emph{};
  \href{https://doi.org/10.1103/PhysRevLett.132.221003}{\emph{Phys. Rev. Lett.}
  {\bfseries 132} (2024) 221003},
  [\href{https://arxiv.org/abs/2211.03395}{{\ttfamily 2211.03395}}].

\bibitem{Kristiano:2024ngc}
J.~Kristiano and J.~Yokoyama\emph{};
  \href{https://arxiv.org/abs/2405.12149}{{\ttfamily 2405.12149}}.

\bibitem{Riotto:2023hoz}
A.~Riotto\emph{};  \href{https://arxiv.org/abs/2301.00599}{{\ttfamily
  2301.00599}}.

\bibitem{Choudhury:2023vuj}
S.~Choudhury, M.~R. Gangopadhyay and M.~Sami\emph{};
  \href{https://arxiv.org/abs/2301.10000}{{\ttfamily 2301.10000}}.

\bibitem{Choudhury:2023jlt}
S.~Choudhury, S.~Panda and M.~Sami\emph{};
  \href{https://doi.org/10.1016/j.physletb.2023.138123}{\emph{Phys. Lett. B}
  {\bfseries 845} (2023) 138123},
  [\href{https://arxiv.org/abs/2302.05655}{{\ttfamily 2302.05655}}].

\bibitem{Firouzjahi:2023ahg}
H.~Firouzjahi and A.~Riotto\emph{};
  \href{https://doi.org/10.1088/1475-7516/2024/02/021}{\emph{JCAP} {\bfseries
  02} (2024) 021}, [\href{https://arxiv.org/abs/2304.07801}{{\ttfamily
  2304.07801}}].

\bibitem{Firouzjahi:2023aum}
H.~Firouzjahi\emph{};
  \href{https://doi.org/10.1088/1475-7516/2023/10/006}{\emph{JCAP} {\bfseries
  10} (2023) 006}, [\href{https://arxiv.org/abs/2303.12025}{{\ttfamily
  2303.12025}}].

\bibitem{Chen:2012ye}
X.~Chen and Y.~Wang\emph{};  \href{https://arxiv.org/abs/1207.0100}{{\ttfamily
  1207.0100}}.

\end{thebibliography}\endgroup

\bibliographystyle{JHEPNoTitle}


\end{document}